\documentclass{article}
\usepackage[utf8]{inputenc}
\usepackage[english]{babel}
\usepackage{graphicx}
\usepackage{amssymb}
\usepackage{amsmath}
\usepackage{physics}
\usepackage{abstract}
\usepackage{bbm}
\usepackage{authblk}
\usepackage{subcaption}
\addto\captionsenglish{}    
\usepackage[margin=1in]{geometry}
\usepackage[sort&compress,numbers]{natbib}
\usepackage{doi}
\usepackage{hyperref}
\hypersetup{colorlinks=true,linkcolor=blue}

\title{Nonlinear optical conductivity of a two-band crystal I.}
\author{D. J. Passos}
\author{G. B. Ventura}
\author{J. M. B. Lopes dos Santos}
\author{J. M. Viana Parente Lopes}
\affil{\small{\textit{Centro de Física das Universidades do Minho e Porto, Departamento de Física e Astronomia, Faculdade de Ciências, Universidade do Porto, 4169-007 Porto, Portugal}}}
\date{}

\begin{document}

\maketitle


\def\Xint#1{\mathchoice
	{\XXint\displaystyle\textstyle{#1}}%
	{\XXint\textstyle\scriptstyle{#1}}%
	{\XXint\scriptstyle\scriptscriptstyle{#1}}%
	{\XXint\scriptscriptstyle\scriptscriptstyle{#1}}%
	\!\int}
\def\XXint#1#2#3{{\setbox0=\hbox{$#1{#2#3}{\int}$}
		\vcenter{\hbox{$#2#3$}}\kern-.5\wd0}}
\def\ddashint{\Xint=}
\def\dashint{\Xint-}

\begin{abstract}
    \small
    The structure of the electronic nonlinear optical conductivity is elucidated in a detailed study of the time-reversal symmetric two-band model. The nonlinear conductivity is decomposed as a sum of contributions related with different regions of the First Brillouin Zone, defined by single or multiphoton resonances. All contributions are written in terms of the same integrals, which contain all information specific to the particular model under study. In this way, ready-to-use formulas are provided that reduce the often tedious calculations of the second and third order optical conductivity to the evaluation of a small set of similar integrals. In the scenario where charge carriers are present prior to optical excitation, Fermi surface contributions must also be considered and are shown to have an universal frequency dependence, tunable by doping. General characteristics are made evident in this type of resonance-based analysis: the existence of step functions that determine the chemical potential dependence of electron-hole symmetric insulators; the determination of the imaginary part by Hilbert transforms, simpler than those of the nonlinear Kramers-Kr\"{o}nig relations; the absence of Drude peaks in the diagonal elements of the second order conductivity, among others. As examples, analytical expressions are derived for the nonlinear conductivities of some simple systems: a very basic model of direct gap semiconductors and the Dirac fermions of monolayer graphene.
\end{abstract}

\section{Introduction}

Early in the history of nonlinear optics, shortly after the discovery of second harmonic generation~\cite{Franken1961GenerationHarmonics} and two-photon absorption~\cite{Kaiser1961Two-photonCaF_2Eu2+}, a general framework was set up to describe the variety of observed optical phenomena. This framework is based on the observation that even for the intense laser light required in these experiments, the electric fields involved are often much weaker than the atomic electric fields that bind the electrons, enabling a perturbative treatment of the light-matter interaction~\cite{Franken1961GenerationHarmonics}. By means of an expansion of the electric current (polarization) in powers of the optical fields, it is possible to define nonlinear optical conductivities (susceptibilities) and describe much of nonlinear optics~\cite{Butcher1991ElementsOptics}. These nonlinear response functions define which effects are present in a given medium, their frequency dependence and their efficiencies. For the purposes of this work, it will be assumed that the electronic contribution to the conductivity is dominant.

Calculations of the nonlinear optical conductivity usually follow a semiclassical approach based on the density matrix formalism. The simplest treatments neglect electron-electron interactions and are within the electric dipole approximation; these assumptions are adopted here as well. For atomic and molecular systems, general expressions for the conductivities are easily derived from perturbation theory and can, by the use of inherent permutation symmetries\footnote{We are referring here to overall permutation symmetry as applied to $\sigma^{\beta\alpha_1\dots\alpha_n}(\bar{\omega}_1,...,\bar{\omega}_n)$ with complex frequencies. See~\cite{Butcher1964TheII} and pages 70-74 of~\cite{Butcher1991ElementsOptics}}, be reduced to a single term~\cite{Butcher1991ElementsOptics}. Despite the apparent simplicity of the nonlinear optical response in these systems, the requirement of knowledge of the exact eigenstates and energies of the unperturbed Hamiltonian $H_0$, together with other complications such as local field effects, makes calculations for any but the simplest molecules less straightforward. Still, semi-empirical ``sum-over-states" calculations have been successful in achieving quantitative agreement with experiment~\cite{Lalama1979OriginSystems,Teng1983DispersionP-Nitroaniline}.

The treatment of the nonlinear optical response of a solid is, however, more complex. Integration over the First Brillouin Zone (FBZ) is necessary and defining the perturbation matrix elements in the Bloch basis can be a nontrivial task. The first full band structure calculations for semiconductors were performed more than a decade after the general expressions for the nonlinear conductivity were derived~\cite{Butcher1963TheI,Butcher1964TheII,Fong1975TheoreticalInSb,Moss1987EmpiricalCrystals,Moss1990Band-structureGaAs}. A first look at the minimal coupling Hamiltonian $\hat{H}_0(\hat{\textbf{r}},\hat{\textbf{p}}+e\,\boldsymbol{A}(t))$ used in early attempts to study nonlinear optics in solids suggests that a crystal could be seen as a collection of independent atomic systems, one for each \textbf{k} in the FBZ. This is indeed true when the entirety of the band structure is taken into account. However, working with an infinite number of bands is hardly practical and when we inevitably settle with an approximate model with a finite number of bands, this description is shown to be inadequate and filled with unphysical infrared divergences~\cite{Sipe1993NonlinearApproximation,Ventura2017GaugeResponses}. Only recently has this problem been resolved with a reformulation of the perturbation theory that extends its validity to systems with finite number of bands~\cite{Passos2018NonlinearAnalysis,Parker2019DiagrammaticSemimetals,Joao2018Non-linearMethods,Ventura2020AApproximation}.

In the nineties, the difficulties associated with the use of the minimal coupling Hamiltonian (also called ``velocity gauge") led to the development of an alternative, if equivalent, perturbation theory, corresponding to a different gauge choice (termed ``length gauge") in representing the optical field~\cite{Aversa1995NonlinearAnalysis}. In this gauge, the Hamiltonian has the form $\hat{H}=\hat{H}_0+e\,\hat{\mathbf{r}}\cdot\boldsymbol{E}(t)$, where $e$ is the electron charge, $\hat{\mathbf{r}}$ is the position operator and $\boldsymbol{E}$ is the classical optical field. Expressions were derived for the second order conductivity of cold semiconductors that are applicable to models with a finite number of bands and no spurious divergences are found in this formulation~\cite{Hughes1996CalculationSemiconductors,Sipe2000Second-orderSemiconductors}. The length gauge approach has since become a well established method of obtaining nonlinear optical response functions~\cite{Aversa1994ThirdModel,Hughes1997CalculationAlN,Cheng2014ThirdGraphene,Cheng2015ThirdTemperature}.

In the length gauge, both intraband and interband transitions must be taken into account and are transparently expressed in the structure of the perturbation theory, which contains, as particular cases, the dynamics of atomic systems and of the free carriers single-band motion, but is more general, and complex, than either~\cite{Aversa1995NonlinearAnalysis}. The complexity stems from expressing the perturbation in terms of a position operator which takes the form of a covariant derivative in the Bloch representation~\cite{Blount1962FormalismsTheory,Aversa1995NonlinearAnalysis,Ventura2017GaugeResponses}. The successive application of derivatives as we move to higher orders in perturbation theory leads to unwieldy expressions for the nonlinear conductivity and lengthy calculations even for the simplest models, the only ones for which analytical calculations are even attempted. Despite this, the results sometimes show surprising simplicity and structure. As an example, the third order conductivity of the system of massless Dirac fermions found in monolayer graphene has the form\footnote{In the relaxation-free limit.}~\cite{Cheng2014ThirdGraphene},

\begin{equation}
    \sigma^{xxxx}(\omega,\omega,\omega)=\frac{C_0}{\omega^4}\left(-17\,G\left(\frac{\hbar\omega}{2|\mu|}\right)+64\,G\left(\frac{2\hbar\omega}{2|\mu|}\right)-45\,G\left(\frac{3\hbar\omega}{2|\mu|}\right)\right)
    \label{diracTHG}
\end{equation}
with

\begin{equation}
    G(x)\equiv\Theta(|x|-1)+\frac{i}{\pi}\log\left|\frac{1-x}{1+x}\right|
\end{equation}
where $\omega$ is the frequency of the incident monochromatic optical field, $\mu$ is the chemical potential, $v_F$ is the Fermi velocity and $C_0$ is a constant: $C_0=v_F^2\,e^4/192\hbar^3$.

In physics, when elaborate and extensive calculations are required to derive simple and elegant results, one is led to suspect that a simpler and more insightful way to express the theory exists. This is the perspective we take here. Eq.~\ref{diracTHG} has the third order conductivity broken up into pieces that are relevant in different regions of the spectrum, depending on whether one-, two- or three-photon frequencies are closer to matching the ``effective gap" given by $2|\mu|$. This suggests that a resonance-based decomposition of the conductivity might be possible in general, leading to a more direct derivation of analytical results and easier interpretation of the underlying physics.

In exploring this possibility, we confine ourselves to the study of the two-band model, the solid-state analogue of the two-level atom. In similar spirit to the theoretical investigations of the nonlinear optics of two-level atoms during the seventies~\cite{Butcher1991ElementsOptics,Allen1987OpticalAtoms}, we expect a study of the two-band model to provide a firm foundation for later investigations of more general systems, to allow the central concepts to emerge more simply and to have a wide range of applicability, encompassing any situation where the incident photon frequencies connect a single pair of conduction $(c)$ and valence $(v)$ bands.

As described in previous work~\cite{Aversa1995NonlinearAnalysis,Ventura2017GaugeResponses,Passos2018NonlinearAnalysis}, for the purpose of understanding nonlinear optical properties, the specification of the band structure $\epsilon_{\mathbf{k}a}$ and the non-abelian Berry connection $\mathbf{\mathcal{A}}_{\mathbf{k}ab}$, with $a$ and $b$ as band indices running over $\{c,v\}$, defines the electronic system under study. For simplicity, it is assumed that the system is time-reversal invariant: $\epsilon_{-\mathbf{k}a}=\epsilon_{\mathbf{k}a}$ and $\mathbf{\mathcal{A}}_{-\mathbf{k}ba}=\mathbf{\mathcal{A}}_{\mathbf{k}ab}$ (for the right choice of phases in the Bloch functions). The generalization to systems with broken time-reversal symmetry will be deferred to a future work.

In the next section, the well-known semiclassical density matrix perturbation theory, formulated in the length gauge, is briefly reviewed. Complex frequencies are used throughout this work and are introduced here. The central results are in Section~\ref{Resonancebasedanalysis}, where the general expressions for the nonlinear conductivity of the two-band system are presented. A decomposition is made based on the possible resonances and ready-to-use formulas are provided that, for any two-band model, reduce the calculation of the second and third order conductivity to the evaluation of two and six integrals over the FBZ, respectively. The integrals can sometimes be evaluated analytically in the relaxation-free limit, as described in Section~\ref{Relaxationfree}. In this limit, integral evaluation is analogous to performing a series of Fermi golden rule calculations, which determine the real part of the nonlinear conductivity, followed by consideration of the respective Hilbert transforms, that provide the imaginary part. For systems that possess a Fermi surface (e.g. metals), there is one additional integral to compute at each order. A brief discussion on how to obtain finite temperature results from a zero temperature calculation is included at the end of the section, closing the exposition of the formalism. The ideas and tools developed in this work are illustrated with calculations for a model with parabolic isotropic bands and constant matrix elements $\mathcal{A}_{\mathbf{k}ab}$ (an overly simplistic description of semiconductors) in Section~\ref{Parabolic} and for the system of massless Dirac fermions present in monolayer graphene in Section~\ref{Monolayergraphene}. The main conclusions are summarized in Section~\ref{Conclusions}.\label{Introduction}

\section{Perturbation theory} \label{Perturbationtheory}

In a system of non-interacting electrons moving through the periodic potential of a crystal, the charge motion that is perturbed by a passing light wave can be appropriately described, if the optical fields $\boldsymbol{E}(t)$ are sufficiently weak, by performing an expansion of the current density $\boldsymbol{J}(t)$ in a power series,

\begin{equation}
    \boldsymbol{J}(t)=\boldsymbol{J}^{(1)}(t)+\boldsymbol{J}^{(2)}(t)+\dots+\boldsymbol{J}^{(n)}(t)+\dots
    \label{expansion}
\end{equation}

The linear term in this expansion provides the usual definition of the optical conductivity,

\begin{equation}
    J^{\beta\,(1)}(t)=\int_{-\infty}^{+\infty}\sigma^{\beta\alpha}(t-t')\,E^{\alpha}(t')\,dt'
    \label{j(t)linear}
\end{equation}
with an implicit summation over repeated tensor indices $\alpha$. Similarly, we can define nonlinear conductivities by taking into account higher powers of $\mathbf{E}$,

\begin{equation}
    J^{\beta\,(n)}(t)=\int_{-\infty}^{+\infty}\dots\int_{-\infty}^{+\infty}\sigma^{\beta\alpha_1\dots\alpha_n}(t-t_1,\dots,t-t_n)\,E^{\alpha_1}(t_1)\dots E^{\alpha_n}(t_n)\,dt_1\dots dt_n
    \label{j(t)}
\end{equation}

This defines the constitutive relation $J(E)$ in the time-domain. Since we shall be concerned with the pole structure of the nonlinear conductivity and resonances are more naturally discussed in the frequency domain, we ought to rewrite the previous equation,

\begin{equation}
    J^{\beta\,(n)}(t)=\int_{-\infty}^{+\infty}\dots\int_{-\infty}^{+\infty}\frac{d\omega_1}{2\pi}\dots \frac{d\omega_n}{2\pi}\,\,\sigma^{\beta\alpha_1\dots\alpha_n}(\omega_1,\dots,\omega_n)\,E^{\alpha_1}(\omega_1)\dots E^{\alpha_n}(\omega_n)\,e^{-i(\omega_1+\dots+\omega_n)t}
    \label{j(omega)}
\end{equation}
with

\begin{equation}
    \sigma^{\beta\alpha_1\dots\alpha_n}(\omega_1,\dots,\omega_n)\equiv\int_{-\infty}^{+\infty}\dots\int_{-\infty}^{+\infty}\sigma^{\beta\alpha_1\dots\alpha_n}(t_1,\dots,t_n)\,e^{i\omega_1t_1}\dots e^{i\omega_nt_n}\,dt_1\dots dt_n
    \label{fouriertransform}
\end{equation}

Due to causality, the time-domain conductivity is nonzero only for positive times: $\sigma(\dots,t_i,\dots)=0$ if $t_i<0$, setting a lower limit in the integration range of Eq.~\ref{fouriertransform} and an upper limit in Eqs.~\ref{j(t)linear} and~\ref{j(t)}. This is seen explicitly in the expressions derived from perturbation theory (Eqs.~\ref{jn} and~\ref{sigmat}).

\subsection{Density matrix formalism}

The dynamics of the current density, when thermally averaged, is given by the density matrix $\hat{\rho}$,

\begin{equation}
    \boldsymbol{J}(t)=\Tr(\boldsymbol{\hat{J}}\,\hat{\rho}(t))
    \label{tr}
\end{equation}
whose time evolution follows the Liouville equation,

\begin{equation}
     i\hbar\,\partial_t\hat{\rho}=[\hat{H},\hat{\rho}]=[\hat{H}_0,\hat{\rho}]+e\left[\hat{r}^{\alpha},\hat{\rho}\right]E^{\alpha}(t)
     \label{liouville}
\end{equation}

To perform the expansion in Eq.~\ref{expansion}, the density matrix itself must be expanded as a powers series in the optical fields. Perturbative solutions to Eq.~\ref{liouville} can then be found and replaced in Eq.~\ref{tr}. When solving Eq.~\ref{liouville}, it is assumed that in the infinite past, when the perturbation was absent, the system was in an equilibrium described by the Fermi-Dirac distribution $\hat{\rho}(t=-\infty)=\hat{\rho}_0\equiv f(\hat{H}_0)$.

At order $n$,

\begin{equation}
    J^{\beta(n)}(t)=\left(-\frac{i\,e}{\hbar}\right)^n\int_{-\infty}^{t}dt_n\dots\int_{-\infty}^{t_2}dt_1\,\Tr\left(\hat{J}^{\beta}\left[\hat{r}^{\alpha_n}_I(t_n-t),\dots\left[\hat{r}^{\alpha_1}_I(t_1-t),\hat{\rho}_0\right]\dots\right]\right)\,E^{\alpha_1}(t_1)\dots E^{\alpha_n}(t_n)
    \label{jn}
\end{equation}
where the subscript $I$ stands for the interaction representation $r_I(t)=e^{iH_0t/\hbar}\,r\,e^{-iH_0t/\hbar}$.

From Eqs.~\ref{j(t)} and~\ref{jn}, we arrive at the nonlinear conductivity,

\begin{equation}
    \sigma^{\beta\alpha_1\dots\alpha_n}(t_1,\dots ,t_n)=\left(-\frac{i\,e}{\hbar}\right)^n\Tr\left(\hat{J}^{\beta}\,\left[\hat{r}^{\alpha_n}_I(-t_n),\dots\left[\hat{r}^{\alpha_1}_I(-t_1),\hat{\rho}_0\right]\dots\right]\right)\,\Theta(t_1-t_2)\dots\Theta(t_{n-1}-t_n)\Theta(t_n)
    \label{sigmat}
\end{equation}

As already mentioned, we are interested in the frequency domain version of Eq.~\ref{sigmat}. It could be derived from Eq.~\ref{sigmat} by a Fourier transform (Eq.~\ref{fouriertransform}) or by solving Eq.~\ref{liouville} directly in the frequency domain. The trace in Eq.~\ref{sigmat} is to be evaluated over the eigenstates of the unperturbed Hamiltonian $H_0$, the Bloch functions $\psi_{\mathbf{k}a}$ with energies $\epsilon_{\mathbf{k}a}$; the trace involves not only a sum over discrete band indices $a$ and $b$ but also an integration over a continuum variable $\mathbf{k}$ that takes any value in the FBZ. The end result is that the nonlinear optical conductivity takes the form,
    
\begin{align}
    \sigma^{\beta\alpha_1\dots\alpha_n}(\omega_1,\dots,\omega_n)=e^n\,\int\frac{d^d\mathbf{k}}{(2\pi)^d}&\sum_{a,b}\frac{J^{\beta}_{\mathbf{k}ba}}{\hbar\omega_1+\dots+\hbar\omega_n-\Delta\epsilon_{\mathbf{k}ab}}\nonumber\\
    \times&\left[r^{\alpha_n},\dots\frac{1}{\hbar\omega_1+\hbar\omega_2-\Delta\epsilon_{\mathbf{k}}}\circ\left[r^{\alpha_2},\frac{1}{\hbar\omega_1-\Delta\epsilon_{\mathbf{k}}}\circ\left[r^{\alpha_1},\rho_0\right]\right]\dots\right]_{\mathbf{k}ab}
    \label{sigmaomega}
\end{align}
with $\circ$ as the Hadamard, or element-wise, product, $d$ as the dimensionality of the system and $\Delta\epsilon_{\mathbf{k}ab}\equiv\epsilon_{\mathbf{k}a}-\epsilon_{\mathbf{k}b}$. A more detailed derivation of Eq.~\ref{sigmaomega} can be found elsewhere~\cite{Ventura2017GaugeResponses}.

Aside from the integration over the FBZ, the expression for the nonlinear conductivity in Eq.~\ref{sigmaomega} is identical to that of atomic systems. The essential differences emerge when attempting to write the position operator in the basis of eigenstates of $H_0$. While in the case of atomic systems $\hat{r}$ has a well-behaved representation, in crystals the situation is analogous to that of a free particle, where a representation based on momentum eigenstates is used. In the momentum basis, the position operator takes the form of a derivative. In the Bloch basis, $\hat{r}$ is the covariant derivative~\cite{Blount1962FormalismsTheory,Ventura2017GaugeResponses},

\begin{equation}
    r^{\alpha}_{\mathbf{k}ab}=i\,D^{\alpha}_{\mathbf{k}ab}\equiv i\left(\delta_{ab}\,\frac{\partial}{\partial k^{\alpha}}-i\,\mathcal{A}^{\alpha}_{\mathbf{k}ab}\right)
    \label{covariantderivative}
\end{equation}

In their pioneering work~\cite{Aversa1995NonlinearAnalysis}, Aversa and Sipe noted that the representation of the position operator should not cause any real difficulties as the operator occurs only in well-defined commutators,

\begin{equation}
    [\hat{r}^{\alpha},\hat{\mathcal{O}}]_{\mathbf{k}ab}=i\,[\hat{D}^{\alpha},\hat{\mathcal{O}}]_{\mathbf{k}ab}=i\,\partial^{\alpha}\mathcal{O}_{\mathbf{k}ab}+[\hat{\mathcal{A}}^{\alpha},\hat{\mathcal{O}}]_{\mathbf{k}ab}
    \label{dcommutator}
\end{equation}
where $\hat{\mathcal{O}}$ stands for a generic observable with well-defined matrix elements, diagonal in $\mathbf{k}$.

In particular, the matrix elements of the current density,

\begin{equation}
    J^{\beta}_{\mathbf{k}ba}=-e\,\frac{(-i)}{\hbar}\,[\hat{r}^{\beta},\hat{H}]_{\mathbf{k}ba}=-\frac{e}{\hbar}\,[\hat{D}^{\beta},\hat{H}_0]_{\mathbf{k}ba}=-\frac{e}{\hbar}\left(\delta_{ab}\,\partial^{\beta}\epsilon_{\mathbf{k}a}-i\,\mathcal{A}^{\beta}_{\mathbf{k}ba}\,\Delta\epsilon_{\mathbf{k}ab}\right)
    \label{currentmatrixelements}
\end{equation}
with the use of the commutation relation $\left[r^{\beta},r^{\alpha}\right]=0$ (Appendix \hyperlink{A}{A}).

\subsection{Complex frequencies} \label{Complexfrequencies}

A careful inspection of the general formula derived in the previous section raises some subtle issues. For any frequency $\omega$ for which a resonance exists, the denominators in Eq.~\ref{sigmaomega} are strictly undefined. This difficulty traces back to Eq.~\ref{fouriertransform} where the existence of a frequency domain nonlinear conductivity relies on the convergence of the Fourier transform. When ignoring any kind of relaxation mechanism, the response to an impulse given at an instant of time can last forever: $\sigma(t=+\infty)\neq 0$ and the Fourier transform diverges. This problem can be circumvented by extending the definition in Eq.~\ref{fouriertransform} to complex frequencies in the upper half-plane~\cite{Butcher1991ElementsOptics},

\begin{align}
    \sigma^{\beta\alpha_1\dots\alpha_n}(\bar{\omega}_1,\dots,\bar{\omega}_n)&=\int_{-\infty}^{+\infty}\dots\int_{-\infty}^{+\infty}\sigma^{\beta\alpha_1\dots\alpha_n}(t_1,\dots,t_n)\,e^{i\bar{\omega}_1t_1}\dots e^{i\bar{\omega}_nt_n}\,dt_1\dots dt_n\nonumber\\
    &=\int_{-\infty}^{+\infty}\dots\int_{-\infty}^{+\infty}\left(\sigma^{\beta\alpha_1\dots\alpha_n}(t_1,\dots,t_n)\,e^{-\gamma t_1}\dots e^{-\gamma t_n}\right)\,e^{i\omega_1t_1}\dots e^{i\omega_nt_n}\,dt_1\dots dt_n
    \label{complexfrequencies}
\end{align}
with $\omega=\Re(\bar{\omega})$ and $\gamma=\Im(\bar{\omega})$.

The extension to complex frequencies can be interpreted in two ways. One is to consider the response function in Eq.~\ref{complexfrequencies} to be associated not to monochromatic waves, but to fields that are adiabatically switched on in the infinite past $E(\omega)\,e^{-i\omega t}\,e^{\gamma t}$. A different perspective is to look at complex frequencies as a simple phenomenological method to introduce relaxation into the system. As stated in Eq.~\ref{complexfrequencies}, the nonlinear conductivity in the (real) frequency domain $\sigma'(\omega_1,\dots,\omega_n)\equiv\sigma(\bar{\omega}_1,\dots,\bar{\omega}_n)$ is obtained from a Fourier transform of a time domain response function that has the form $\sigma'(t_1,\dots,t_n)=\sigma(t_1,\dots,t_n)\,e^{-\gamma t_1}\dots e^{-\gamma t_n}$ and satisfies the condition $\sigma'(t_i=\infty)=0$. This approach to relaxation is most certainly too simplistic to properly account for all the possible relaxation mechanisms that are observed in experiments, but it provides a direct and easy way to obtain sensible answers and it has advantages relative to the traditional approach of adding an phenomenological term to the equation of motion (Eq.~\ref{liouville})~\cite{Passos2018NonlinearAnalysis,Holder2020ConsequencesMotion}. For simplicity, we here take the imaginary part of the frequencies to be a constant $\gamma$ but more generally we could have $\gamma=\gamma(\omega)$. It would only be required that the function be even, in order for the reality condition to be maintained.

When the relaxation-free limit is considered, where the imaginary parts of the frequencies are taken to zero from above, the integrand in Eq.~\ref{sigmaomega} can always be defined as a distribution by making use of the Sokhotski-Plemelj theorem,

\begin{equation}
    \int\frac{d^d\mathbf{k}}{(2\pi)^d}\,\frac{g_{\mathbf{k}}}{\hbar\bar{\omega}-\Delta\epsilon_{\mathbf{k}}}\xrightarrow{\gamma\rightarrow 0^+}\dashint\frac{d^d\mathbf{k}}{(2\pi)^d}\frac{g_{\mathbf{k}}}{\hbar\omega-\Delta\epsilon_{\mathbf{k}}}-i\,\pi\int\frac{d^d\mathbf{k}}{(2\pi)^d}\,g_{\mathbf{k}}\,\delta(\hbar\omega-\Delta\epsilon_{\mathbf{k}})
    \label{sokhotskiplemelj}
\end{equation}

For an atomic system, the distribution will be defined relative to an integral over the frequencies (Eq.~\ref{j(omega)}) and relies on the condition that the spectral width of $E(\omega)$ be much greater than $\gamma$. For a crystal, the distribution is accommodated by the presence of an integration over the FBZ and no restrictions must be applied to the optical fields considered. By taking the limit $\gamma\rightarrow 0^+$ in the expression

\begin{equation}
    \sigma^{\beta\alpha_1\dots\alpha_n}(\bar{\omega}_1,\dots,\bar{\omega}_n)=e^n\,\int\frac{d^d\mathbf{k}}{(2\pi)^d}\sum_{a,b}    \frac{J^{\beta}_{\mathbf{k}ba}}{\hbar\bar{\omega}_1+\dots+\hbar\bar{\omega}_n-\Delta\epsilon_{\mathbf{k}ab}}\,\left[r^{\alpha_n},\dots\frac{1}{\hbar\bar{\omega}_1-\Delta\epsilon_{\mathbf{k}}}\circ\left[r^{\alpha_1},\rho_0\right]\dots\right]_{\mathbf{k}ab}
    \label{sigmaomegabar}
\end{equation}
the nonlinear conductivity can be defined as a regular function for real frequencies and vanishing relaxation. It is implicitly assumed in this reasoning that no more than a single denominator diverges at a given $\mathbf{k}$ (see Section~\ref{Relaxationfree}).

In Eq.~\ref{sigmaomegabar}, the nonlinear conductivity can be further extended into the lower half-plane by analytic continuation~\cite{Butcher1964TheII}. In this way, Eq.~\ref{sigmaomegabar} provides a valid expression over the entire complex frequency plane, even in regions where Eq.~\ref{fouriertransform} no longer applies and the response function is, therefore, no longer physical.

\section{Resonance-based analysis} \label{Resonancebasedanalysis}

From Eq.~\ref{sigmaomegabar}, a more explicit form of the nonlinear conductivity can be derived by expanding out all the commutators and performing all the required differentiations (that follow from Eq.~\ref{dcommutator}), resulting in a lengthy and rather cumbersome formula. This is the usual starting point when computing the nonlinear optical response functions of semiconductors and other materials. Numerical integration is necessary except for some cases where a low-energy description exists with very simple dispersion relation and eigenstates. For these systems, analytical calculations are sometimes possible but still often rather long and complicated. In this section, we attempt to bring some simplicity and clarity to the structure of the nonlinear conductivity, by separating out terms whose resonances are located in different regions of the FBZ.

Since we are restricting ourselves to the analysis of a two-band system, there is a single (nonzero) energy difference in the denominators of Eq.~\ref{sigmaomegabar}, $\Delta\epsilon_{ab}=\pm\Delta\epsilon_{cv}$, allowing for a partial fraction decomposition into terms with a single denominator to be integrated, $(\hbar\bar{\omega}_1+\dots+\hbar\bar{\omega}_i\pm\Delta\epsilon_{cv})^{-1}$ with $i\in\{1,\dots,n\}$. These terms we denote by $\sigma^{\beta\alpha_1\dots\alpha_n}_i(\bar{\omega}_1,\dots,\bar{\omega}_i,\dots,\bar{\omega}_n)$ as they are associated with resonances involving an $i$ number of photons\footnote{In due rigor, we work with a classical electromagnetic field and there are no photons present. It should however be clear that when a proper quantum treatment is made $\hbar\omega_i$ is the energy of an incident photon, thereby justifying the used nomenclature.}. We shall see later how the real part of these contributions is entirely described by the properties of the crystal in the vicinity of regions of the FBZ where the resonance condition $\hbar\omega_1+\dots+\hbar\omega_i-\Delta\epsilon_{cv}=0$ is met. Some terms will involve poles of higher orders, but these can reduced back to simple poles with an integration by parts or, equivalently, by making use of the identities in Appendix \hyperlink{B}{B}. Finally, there will be terms where the application of the position operator resulted in derivatives of the Fermi-Dirac distribution. These terms will be treated separately and are denoted by $\sigma_F^{\beta\alpha_1\dots\alpha_n}(\bar{\omega}_1,\dots,\bar{\omega}_n)$. An explicit application of the procedure outlined here can be found in Appendix \hyperlink{C}{C}, where the second order conductivity is treated in detail. For now, we state it generally,

\begin{equation}
    \sigma^{\beta\alpha_1\dots\alpha_n}(\bar{\omega}_1,\dots,\bar{\omega}_n)=\sigma_F^{\beta\alpha_1\dots\alpha_n}(\bar{\omega}_1,\dots,\bar{\omega}_n)+\sigma_1^{\beta\alpha_1\dots\alpha_n}(\bar{\omega}_1,\dots,\bar{\omega}_n)+\cdots+\sigma^{\beta\alpha_1\dots\alpha_n}_n(\bar{\omega}_1,\dots,\bar{\omega}_n)
    \label{resonancebasedanalysis}
\end{equation}

The various pieces of Eq.~\ref{resonancebasedanalysis} will be made explicit in the following sections, but it is useful to first inspect their structure in general terms. The one-photon contribution can always be written as

\begin{equation}
    \sigma_1^{\alpha}(\bar{\omega}_1,\cdots,\bar{\omega}_n)=\sum_j\sum_p C_{1 j}^{p}(\bar{\omega}_1,\cdots,\bar{\omega}_n)\,\Pi_j^{p(\alpha)}(\bar{\omega}_1)
    \label{onephoton}
\end{equation}
where all tensor indices where condensed into one $\alpha\equiv\beta\alpha_1\dots\alpha_n$ and $p$ stands for permutation. The sum in $p$ implies $p(\alpha)$ runs over all permutations of $\alpha$, with a specific coefficient for each permutation applied. The coefficients $C_{1j}^{p}(\bar{\omega}_1,\cdots,\bar{\omega}_n)$ are specified in the following sections for the linear, second order and third order conductivities ($n=1$, $2$ and $3$, respectively), where it is observed that most of these coefficients are zero, making only a small number of permutations necessary in practice. The coefficients are independent of the details of the system under consideration (they depend solely on the optical frequencies). All dependence on material properties in the sum of Eq.~\ref{onephoton} is in the integrals $\Pi_{j}^{\alpha}$ that take the general form,

\begin{equation}
    \Pi_j^{\alpha}(\bar{\omega})=\int\frac{d^d\mathbf{k}}{(2\pi)^d}\sum_{a,b}\frac{g_j^{\alpha}(\mathcal{A},\epsilon)_{ab}}{\hbar\bar{\omega}-\Delta\epsilon_{ab}}\,\Delta f_{ba}
    \label{piintegrals}
\end{equation}
with $\Delta f_{ba}\equiv f(\epsilon_b)-f(\epsilon_a)$ as a difference in Fermi functions and $g_j^{\alpha}$ as a set of functions, labeled by $j=1,2,\dots$, that depend on the energies and their derivatives and on the non-abelian Berry connection $\mathcal{A}$ and its generalized derivatives. For notational ease, the $\mathbf{k}$ label has been dropped.

Similarly, for the two-photon contributions,

\begin{equation}
    \sigma_2^{\alpha}(\bar{\omega}_1,\cdots,\bar{\omega}_n)=\sum_j\sum_p C_{2 j}^{p}(\bar{\omega}_1,\cdots,\bar{\omega}_n)\,\Pi_j^{p(\alpha)}(\bar{\omega}_1+\bar{\omega}_2)
    \label{twophoton}
\end{equation}
and the generalization is obvious at this point,

\begin{equation}
    \sigma_i^{\alpha}(\bar{\omega}_1,\cdots,\bar{\omega}_n)=\sum_j\sum_p C_{i j}^{p}(\bar{\omega}_1,\cdots,\bar{\omega}_n)\,\Pi_j^{p(\alpha)}(\bar{\omega}_1+\cdots+\bar{\omega}_i)
    \label{iphoton}
\end{equation}

Since all contributions involve a combination of the same integrals with a changing argument, the calculation of the nonlinear conductivity is reduced to the evaluation of the integrals in Eq.~\ref{piintegrals}. The complexity and number of integrals to be evaluated increases with the order $n$ of the nonlinear conductivity, but they always retain the general form of Eq.~\ref{piintegrals} with varying $g$ functions. For finite $\gamma$, numerical integration will invariably be required. In Section~\ref{Relaxationfree}, we analyze the limit of vanishing relaxation where analytical results are accessible.

It is worth noting at this point that the conductivity in Eq.~\ref{resonancebasedanalysis} is not symmetrized. It follows from the definition in Eq.~\ref{j(omega)} that only the portion of the nonlinear conductivity that respects intrinsic permutation symmetry is physical~\cite{Butcher1991ElementsOptics}. When permutations of Eq.~\ref{resonancebasedanalysis} are properly accounted for, there will not only be an one-photon contribution associated to the resonance $\hbar\omega_1=\Delta\epsilon_{cv}$ but also $\hbar\omega_2=\Delta\epsilon_{cv}$ and so on. Likewise for contributions associated with higher numbers of photons. Having this in mind, the formulas presented for the nonlinear conductivity in this work will nonetheless be left, for the most part, unsymmmetrized. Symmetrizing the conductivity is a trivial, if cumbersome, procedure and adds little to the discussion here.

The terms described so far give a complete description for any two-band model that is used to describe an insulator or a cold semiconductor. But for systems with charge carriers on the Fermi surface prior to optical excitation, there is an additional contribution\footnote{The validity of Eq.~\ref{fermi} has been checked by the authors only up to third order.},

\begin{equation}
    \sigma_F^{\alpha}(\bar{\omega}_1,\cdots,\bar{\omega}_n)=\sum_{X={A,B,...}}\sum_{p}C_X^{p}(\bar{\omega}_1,\cdots,\bar{\omega}_n)\,F_X^{p(\alpha)}
    \label{fermi}
\end{equation}
where the integrals have a different structure than before,

\begin{equation}
    F_X^{\beta\alpha_1\cdots\alpha_n}=\int\frac{d^d\mathbf{k}}{(2\pi)^d}\sum_{a}g_X^{\alpha_1\dots\alpha_n}(\mathcal{A},\epsilon)_{aa}\,\,\partial^{\beta}f_a
    \label{fermiintegrals}
\end{equation}

It is evident from the presence of derivatives of Fermi functions in Eq.~\ref{fermiintegrals} that these integrals are determined by the properties of the Fermi surface. More surprising is the absence of any frequency dependence. All dispersion in $\sigma_F$ comes from the $C$ coefficients, which are the same for every two-band system. This leads us to an important result: \textit{the dispersion of the contributions from the Fermi surface}, those that dominate the optical response of metals at low frequencies, \textit{is given by an universal family of functions of frequency} (Eq.~\ref{fermi}), obtained through linear combinations of the $C_X$'s. The particular linear combination observed is set by the integrals $F_X$. Being dictated by Fermi surface properties, they are particularly dependent on carrier concentration and can therefore be tuned by doping.

In the following sections, Eqs.~\ref{onephoton}-\ref{fermiintegrals} are made explicit for the linear, second and third order conductivities. Ready-to-use formulas are presented that reduce the calculation of the nonlinear conductivities to the evaluation of a minimal number of integrals over the FBZ (Eqs.~\ref{FAfirst}-\ref{Pi1first},~\ref{FBsecond}-\ref{Pi2second},~\ref{FAthird}-\ref{Pi6third}).

\subsection{Linear order}\label{Linearorder}

In linear order, the resonance-based decomposition of Eq.~\ref{resonancebasedanalysis} falls into the familiar intra- and interband separation,

\begin{equation}
    \sigma^{\beta\alpha_1}(\bar{\omega}_1)=\sigma_F^{\beta\alpha_1}(\bar{\omega}_1)+\sigma_1^{\beta\alpha_1}(\bar{\omega}_1)
    \label{sigma1}
\end{equation}
with

\begin{align}
    \sigma_F^{\beta\alpha_1}(\bar{\omega}_1)&=-\frac{i e^2}{\hbar}\,\frac{1}{\hbar\bar{\omega}_1}\,F_A^{\beta\alpha_1}\label{sigma1F}\\
    \sigma_1^{\beta\alpha_1}(\bar{\omega}_1)&=\frac{i e^2}{\hbar}\,\hbar\bar{\omega}_1\,\Pi_1^{\beta\alpha_1}(\bar{\omega}_1)\label{sigma11}
\end{align}

The integrals required to obtain the linear response are well-known and particularly simple. There is one integral associated with an interband resonance and one related to the Fermi surface,

\begin{align}
    F_A^{\beta\alpha_1}&\equiv\int\frac{d^d\mathbf{k}}{(2\pi)^d}\sum_a \partial^{\alpha_1}\epsilon_a\,\partial^{\beta}f_a\label{FAfirst}\\
    \Pi_1^{\beta\alpha_1}(\bar{\omega})&\equiv\int\frac{d^d\mathbf{k}}{(2\pi)^d}\sum_{a,b}\frac{\Re\{\mathcal{A}_{ba}^{\beta}\,\mathcal{A}_{ab}^{\alpha_1}\}}{\hbar\bar{\omega}-\Delta\epsilon_{ab}}\,\Delta f_{ba}\label{Pi1first}
\end{align}\newline
where the real and imaginary parts of products of Hermitian operators are symmetric and antisymmetric, respectively, under exchange of the band labels $a$ and $b$,

\begin{align}
    \Re\{A_{ba}\,B_{ab}\}&=\frac{1}{2}\left(A_{ba} B_{ab}+A_{ab} B_{ba}\right)\label{symmetriccomb}\\
    i\,\Im\{A_{ba}\,B_{ab}\}&=\frac{1}{2}\left(A_{ba} B_{ab}-A_{ab} B_{ba}\right)\label{antisymmetriccomb}
\end{align}

\subsection{Second order}\label{Secondorder}

In second order, Eq.~\ref{resonancebasedanalysis} gives

\begin{equation}
    \sigma^{\beta\alpha_1\alpha_2}(\bar{\omega}_1,\bar{\omega}_2)=\sigma_F^{\beta\alpha_1\alpha_2}(\bar{\omega}_1,\bar{\omega}_2)+\sigma_1^{\beta\alpha_1\alpha_2}(\bar{\omega}_1,\bar{\omega}_2)+\sigma_2^{\beta\alpha_1\alpha_2}(\bar{\omega}_1,\bar{\omega}_2)
    \label{sigma2}
\end{equation}

We now inspect each contribution separately. In the following, we make explicit the $C$ coefficients in Eq.~\ref{iphoton} and the $g$ functions in Eqs.~\ref{piintegrals} and~\ref{fermiintegrals} for $n=2$.

Starting with the Fermi surface contributions,
\begin{align}
    \sigma_F^{\beta\alpha_1\alpha_2}(\bar{\omega}_1,\bar{\omega}_2)=\frac{i e^3}{\hbar}\,\frac{1}{\hbar\bar{\omega}_1}\,F_B^{\alpha_1\alpha_2\beta}
    \label{sigma2F}
\end{align}

The one-photon contributions,

\begin{align}
    \frac{\hbar}{i\,e^3}\,\sigma_1^{\beta\alpha_1\alpha_2}(\bar{\omega}_1,\bar{\omega}_2)=\frac{1}{\hbar\bar{\omega}_1+\hbar\bar{\omega}_2}\,\Pi_1^{\alpha_2\alpha_1\beta}(\bar{\omega}_1)+\frac{\hbar\bar{\omega}_1+\hbar\bar{\omega}_2}{(\hbar\bar{\omega}_2)^2}\,\Pi_1^{\beta\alpha_1\alpha_2}(\bar{\omega}_1)-\frac{\hbar\bar{\omega}_1}{\hbar\bar{\omega}_2}\,\Pi_2^{\alpha_1\beta\alpha_2}(\bar{\omega}_1)
    \label{sigma21}
\end{align}

Finally, the two-photon contributions,

\begin{align}
    \frac{\hbar}{i\,e^3}\,\sigma_2^{\beta\alpha_1\alpha_2}(\bar{\omega}_1,\bar{\omega}_2)=-\frac{\hbar\bar{\omega}_1+\hbar\bar{\omega}_2}{(\hbar\bar{\omega}_2)^2}\,\Pi^{\beta\alpha_1\alpha_2}_1(\bar{\omega}_1+\bar{\omega}_2)+\frac{\hbar\bar{\omega}_1+\hbar\bar{\omega}_2}{\hbar\bar{\omega}_2}\,\Pi^{\beta\alpha_1\alpha_2}_2(\bar{\omega}_1+\bar{\omega}_2)
    \label{sigma22}
\end{align}

In second order, there are two integrals from interband resonances and one from the Fermi surface,

\begin{align}
    F_B^{\beta\alpha_1\alpha_2}&\equiv\int\frac{d^d\mathbf{k}}{(2\pi)^d}\sum_{a}\,\mathcal{F}^{\alpha_1\alpha_2}_a\,\partial^{\beta}f_a\label{FBsecond}\\
    \Pi_1^{\beta\alpha_1\alpha_2}(\bar{\omega})&\equiv\int\frac{d^d\mathbf{k}}{(2\pi)^d}\sum_{a,b}\frac{\Im\{\mathcal{A}^{\beta}_{ba}\,\mathcal{A}^{\alpha_1}_{ab}\}\,(\partial^{\alpha_2}\Delta\epsilon_{ab})}{\hbar\bar{\omega}-\Delta\epsilon_{ab}}\,\Delta f_{ba}\label{Pi1second}\\
    \Pi_2^{\beta\alpha_1\alpha_2}(\bar{\omega})&\equiv\int\frac{d^d\mathbf{k}}{(2\pi)^d}\sum_{a,b}\frac{\Im\{\mathcal{A}^{\beta}_{ba}\,\mathcal{A}^{\alpha_1}_{ab;\alpha_2}\}}{\hbar\bar{\omega}-\Delta\epsilon_{ab}}\,\Delta f_{ba}\label{Pi2second}
\end{align}\newline
where we made use of the generalized derivative notation introduced by Aversa and Sipe~\cite{Aversa1995NonlinearAnalysis}: $\mathcal{O}_{ab;\alpha}\equiv\left(\delta_{ab}\,\partial^{\alpha}-i(\mathcal{A}^{\alpha}_{aa}-\mathcal{A}^{\alpha}_{bb})\right)\mathcal{O}_{ab}$.

We chose, in Eq.~\ref{FBsecond}, to label the integral over the Fermi surface with $B$ instead of $A$ to reserve the notation $F_A$ to the purely intraband contribution. That is, $F_A$ is to be identified as the contribution that would be present even in a system where all interband transitions are neglected. It is however absent in the second order conductivity, due to time-reversal symmetry. It can easily be shown that the same is true for all even orders of the nonlinear conductivity.

While the purely intraband contributions can be derived to any order without the need for a perturbation theory as complex as the one discussed here, the inclusion of interband transitions brings new Fermi surface contributions when we move past linear order. In $F_B^{\beta\alpha_1\alpha_2}$ in particular, we see the appearance of the quantity $\mathcal{F}^{\beta\alpha_1}_{a}\equiv\partial^{\beta}\mathcal{A}^{\alpha_1}_{aa}-\partial^{\alpha_1}\mathcal{A}^{\beta}_{aa}$ which we recognize as the Berry curvature~\cite{Berry1984QuantalChanges} and can be expressed as products of the non-abelian Berry connection (Appendix \hyperlink{A}{A}). It follows that the $\sigma_F$ contribution at second order is a probe of the Berry curvature around the Fermi surface.

Another curious result that follows from Eq.~\ref{FBsecond} is that we can expect the Fermi surface contributions to be absent when the current is measured in the direction of the optical fields: $\sigma_F^{\beta\alpha_1\alpha_2}=0$ when $\beta=\alpha_1=\alpha_2$. As a consequence, \textit{no Drude peaks should be found in the diagonal tensor elements of the second order conductivity.}

Finally, we note that the contributions that probe the FBZ beyond the Fermi surface, the one- and two-photon contributions, are determined by the integrals in Eqs.~\ref{Pi1second} and~\ref{Pi2second} which are known in the literature of the nonlinear optics of solids for their relation to the injection and shift currents in semiconductors~\cite{Sipe2000Second-orderSemiconductors}, respectively.

\subsection{Third order}\label{Thirdorder}

A considerable jump in complexity occurs when we move to third order. Once again, we start with Eq.~\ref{resonancebasedanalysis},

\begin{equation}
    \sigma^{\beta\alpha_1\alpha_2\alpha_3}(\bar{\omega}_1,\bar{\omega}_2,\bar{\omega}_3)=\sigma_F^{\beta\alpha_1\alpha_2\alpha_3}(\bar{\omega}_1,\bar{\omega}_2,\bar{\omega}_3)+\sigma_1^{\beta\alpha_1\alpha_2\alpha_3}(\bar{\omega}_1,\bar{\omega}_2,\bar{\omega}_3)+\sigma_2^{\beta\alpha_1\alpha_2\alpha_3}(\bar{\omega}_1,\bar{\omega}_2,\bar{\omega}_3)+\sigma_3^{\beta\alpha_1\alpha_2\alpha_3}(\bar{\omega}_1,\bar{\omega}_2,\bar{\omega}_3)
    \label{sigma3}
\end{equation}
and write out all contributions explicitly.

The Fermi surface contributions,
\begin{equation}
    \sigma_F^{\beta\alpha_1\alpha_2\alpha_3}(\bar{\omega}_1,\bar{\omega}_2,\bar{\omega}_3)=\frac{i e^4}{6\,\hbar}\,\frac{1}{\hbar\bar{\omega}_1\,\hbar\bar{\omega}_2\,\hbar\bar{\omega}_3}\,F_A^{\beta\alpha_1\alpha_2\alpha_3}
    \label{sigma3F}
\end{equation}

The one-photon contributions,

\begin{align}
    \frac{\hbar}{i\,e^4}\,&\sigma_1^{\beta\alpha_1\alpha_2\alpha_3}(\bar{\omega}_1,\bar{\omega}_2,\bar{\omega}_3)=\nonumber\\
    &-\frac{\hbar\bar{\omega}_{23}}{2\,\hbar\bar{\omega}_2\,\hbar\bar{\omega}_3\,\hbar\bar{\omega}_{123}}\,\Pi_1^{\alpha_1\alpha_2\alpha_3\beta}(\bar{\omega}_1)-\frac{\hbar\bar{\omega}_{13}}{\hbar\bar{\omega}_2\,(\hbar\bar{\omega}_3)^2}\,\Pi_1^{\alpha_1\alpha_2\beta\alpha_3}(\bar{\omega}_1)-\frac{1}{\hbar\bar{\omega}_{12}\,\hbar\bar{\omega}_3}\,\Pi_2^{\alpha_1\alpha_2\alpha_3\beta}(\bar{\omega}_1)\nonumber\\
    &-\frac{\hbar\bar{\omega}_{123}}{2\,\hbar\bar{\omega}_2\,\hbar\bar{\omega}_3\,\hbar\bar{\omega}_{23}}\,\Pi_2^{\beta\alpha_1\alpha_2\alpha_3}(\bar{\omega}_1)-\frac{1}{(\hbar\bar{\omega}_3)^2\,\hbar\bar{\omega}_{123}}\,\Pi_3^{\alpha_1\alpha_2\alpha_3\beta}(\bar{\omega}_1)-\frac{\hbar\bar{\omega}_{123}}{2\,(\hbar\bar{\omega}_2)^2\,(\hbar\bar{\omega}_3)^2}\,\Pi_3^{\beta\alpha_1\alpha_2\alpha_3}(\bar{\omega}_1)\nonumber\\
    &-\frac{\hbar\bar{\omega}_1}{2\,\hbar\bar{\omega}_2\,\hbar\bar{\omega}_3}\Pi_4^{\alpha_1\alpha_2\alpha_3\beta}(\bar{\omega}_1)+\frac{2\,\hbar\bar{\omega}_1\,\hbar\bar{\omega}_{123}}{\hbar\bar{\omega}_2\,\hbar\bar{\omega}_{13}\,\hbar\bar{\omega}_{23}}\,\Pi_5^{\alpha_1\alpha_2\alpha_3\beta}(\bar{\omega}_1)-\frac{(\hbar\bar{\omega}_1)^2\,\hbar\bar{\omega}_{123}}{\hbar\bar{\omega}_2\,\hbar\bar{\omega}_3\,\hbar\bar{\omega}_{12}\,\hbar\bar{\omega}_{13}}\,\Pi_5^{\alpha_2\alpha_3\alpha_1\beta}(\bar{\omega}_1)
    \label{sigma31}
\end{align}
with the abbreviations $\bar{\omega}_{ij}\equiv\bar{\omega}_i+\bar{\omega}_j$ and $\bar{\omega}_{123}\equiv\bar{\omega}_1+\bar{\omega}_2+\bar{\omega}_3$.

The two-photon contributions,

\begin{align}
    \frac{\hbar}{i\,e^4}\,&\sigma_2^{\beta\alpha_1\alpha_2\alpha_3}(\bar{\omega}_1,\bar{\omega}_2,\bar{\omega}_3)=\nonumber\\
    &+\frac{\hbar\bar{\omega}_{12}}{(\hbar\bar{\omega}_2)^2\,\hbar\bar{\omega}_3}\,\Pi_1^{\alpha_1\alpha_3\beta\alpha_2}(\bar{\omega}_{12})-\frac{\hbar\bar{\omega}_{12}}{2\,\hbar\bar{\omega}_1\,\hbar\bar{\omega}_2\,\hbar\bar{\omega}_{123}}\,\Pi_1^{\alpha_3\alpha_1\alpha_2\beta}(\bar{\omega}_{12})-\frac{\hbar\bar{\omega}_{12}\,\hbar\bar{\omega}_{123}}{2\,\hbar\bar{\omega}_1\,\hbar\bar{\omega}_2\,(\hbar\bar{\omega}_3)^2}\,\Pi_1^{\beta\alpha_1\alpha_2\alpha_3}(\bar{\omega}_{12})\nonumber\\
    &+\frac{1}{(\hbar\bar{\omega}_1)^2\,\hbar\bar{\omega}_{123}}\Pi_3^{\alpha_2\alpha_3\alpha_1\beta}(\bar{\omega}_{12})+\frac{\hbar\bar{\omega}_{123}}{(\hbar\bar{\omega}_2)^2\,(\hbar\bar{\omega}_3)^2}\,\Pi_3^{\beta\alpha_1\alpha_2\alpha_3}(\bar{\omega}_{12})-\frac{(\hbar\bar{\omega}_{12})^2}{2\,\hbar\bar{\omega}_1\,\hbar\bar{\omega}_2\,\hbar\bar{\omega}_3}\,\Pi_6^{\alpha_1\alpha_2\alpha_3\beta}(\bar{\omega}_{12})
    \label{sigma32}
\end{align}

Finally, the three-photon contributions,

\begin{align}
   \frac{\hbar}{i\,e^4}\,&\sigma_3^{\beta\alpha_1\alpha_2\alpha_3}(\bar{\omega}_1,\bar{\omega}_2,\bar{\omega}_3)=+\frac{\hbar\bar{\omega}_{12}\,\hbar\bar{\omega}_{123}}{2\,\hbar\bar{\omega}_1\,\hbar\bar{\omega}_2\,(\hbar\bar{\omega}_3)^2}\,\Pi_1^{\beta\alpha_1\alpha_2\alpha_3}(\bar{\omega}_{123})+\frac{\hbar\bar{\omega}_{123}}{2\,\hbar\bar{\omega}_2\,\hbar\bar{\omega}_3\,\hbar\bar{\omega}_{23}}\,\Pi_2^{\beta\alpha_1\alpha_2\alpha_3}(\bar{\omega}_{123})\nonumber\\
   &-\frac{\hbar\bar{\omega}_{123}}{2\,(\hbar\bar{\omega}_2)^2\,(\hbar\bar{\omega}_3)^2}\,\Pi_3^{\beta\alpha_1\alpha_2\alpha_3}(\bar{\omega}_{123})-\frac{\hbar\bar{\omega}_{123}}{\hbar\bar{\omega}_3\,\hbar\bar{\omega}_{23}}\,\Pi_4^{\beta\alpha_1\alpha_2\alpha_3}(\bar{\omega}_{123})-\frac{\hbar\bar{\omega}_{123}}{\hbar\bar{\omega}_{13}\,\hbar\bar{\omega}_{23}}\,\Pi_5^{\alpha_1\alpha_2\alpha_3\beta}(\bar{\omega}_{123})
   \label{sigma33}
\end{align}

In third order, there are six integrals from interband resonances and one from the Fermi surface,

\begin{align}
    F_A^{\beta\alpha_1\alpha_2\alpha_3}&\equiv\int\frac{d^d\mathbf{k}}{(2\pi)^d}\sum_a\,\partial^{\alpha_1}\partial^{\alpha_2}\partial^{\alpha_3}\epsilon_a\,\partial^{\beta}f_a\label{FAthird}\\
    \Pi_1^{\beta\alpha_1\alpha_2\alpha_3}(\bar{\omega})&\equiv\int\frac{d^d\mathbf{k}}{(2\pi)^d}\sum_{a,b}\frac{\Re\{\mathcal{A}^{\beta}_{ba}\,\mathcal{A}^{\alpha_1}_{ab;\alpha_2}\}\,(\partial^{\alpha_3}\Delta\epsilon_{ab})}{\hbar\bar{\omega}-\Delta\epsilon_{ab}}\,\Delta f_{ba}\label{Pi1third}\\
    \Pi_2^{\beta\alpha_1\alpha_2\alpha_3}(\bar{\omega})&\equiv\int\frac{d^d\mathbf{k}}{(2\pi)^d}\sum_{a,b}\frac{\Re\{\mathcal{A}^{\beta}_{ba}\,\mathcal{A}^{\alpha_1}_{ab}\}\,(\partial^{\alpha_2}\partial^{\alpha_3}\Delta\epsilon_{ab})}{\hbar\bar{\omega}-\Delta\epsilon_{ab}}\,\Delta f_{ba}\label{Pi2third}\\
    \Pi_3^{\beta\alpha_1\alpha_2\alpha_3}(\bar{\omega})&\equiv\int\frac{d^d\mathbf{k}}{(2\pi)^d}\sum_{a,b}\frac{\Re\{\mathcal{A}^{\beta}_{ba}\,\mathcal{A}^{\alpha_1}_{ab}\}\,(\partial^{\alpha_2}\Delta\epsilon_{ab})\,(\partial^{\alpha_3}\Delta\epsilon_{ab})}{\hbar\bar{\omega}-\Delta\epsilon_{ab}}\,\Delta f_{ba}\label{Pi3third}\\
    \Pi_4^{\beta\alpha_1\alpha_2\alpha_3}(\bar{\omega})&\equiv\int\frac{d^d\mathbf{k}}{(2\pi)^d}\sum_{a,b}\frac{\Re\{\mathcal{A}^{\beta}_{ba}\,\mathcal{A}^{\alpha_1}_{ab;\alpha_2\alpha_3}\}}{\hbar\bar{\omega}-\Delta\epsilon_{ab}}\,\Delta f_{ba}\label{Pi4third}\\
    \Pi_5^{\beta\alpha_1\alpha_2\alpha_3}(\bar{\omega})&\equiv\int\frac{d^d\mathbf{k}}{(2\pi)^d}\sum_{a,b}\frac{\Re\{\mathcal{A}^{\beta}_{ba}\,\mathcal{A}^{\alpha_1}_{ba}\,\mathcal{A}^{\alpha_2}_{ab}\,\mathcal{A}^{\alpha_3}_{ab}\}}{\hbar\bar{\omega}-\Delta\epsilon_{ab}}\,\Delta f_{ba}\label{Pi5third}\\
    \Pi_6^{\beta\alpha_1\alpha_2\alpha_3}(\bar{\omega})&\equiv\int\frac{d^d\mathbf{k}}{(2\pi)^d}\sum_{a,b}\frac{\Re\{\mathcal{A}^{\beta}_{ba;\alpha_1}\,\mathcal{A}^{\alpha_2}_{ab;\alpha_3}\}}{\hbar\bar{\omega}-\Delta\epsilon_{ab}}\,\Delta f_{ba}\label{Pi6third}
\end{align}\newline

In these equations, all products of off-diagonal matrix elements of $\mathcal{A}$ appear in conjugate pairs (one $\mathcal{A}_{cv}$ and one $\mathcal{A}_{vc}$). This is required by gauge invariance: the nonlinear conductivity must be invariant under a change of phases of the Bloch functions $\psi_{\mathbf{k}a}\rightarrow e^{i\theta_{\mathbf{k}a}}\psi_{\mathbf{k}a}$. For the same reason, generalized instead of regular derivatives of $\mathcal{A}$ must be used.

The numerators of the $\Pi^{\alpha}_j$ integrals involve symmetric (Eq.~\ref{symmetriccomb}) or antisymmetric (Eq.~\ref{antisymmetriccomb}) combinations of a product. Symmetric combinations occur for even orders of perturbation theory and antisymmetric ones for odd orders. This structure is obtained by combining the contributions to the integrands from $\mathbf{k}$ and $-\mathbf{k}$, then invoking time-reversal symmetry (see Appendix \hyperlink{C}{C} for an example). This is an insightful and sometimes useful way to write the integrals. For instance, in evaluating diagonal tensor elements of the second order conductivity it is clear that the integral $\Pi_1$ need not be considered, since it vanishes identically: $\Im\{\mathcal{A}^{\beta}_{ba}\,\mathcal{A}^{\alpha_1}_{ab}\}=0$ for $\beta=\alpha_1$.

The type of resonance-based analysis presented here could be continued for successively higher order nonlinear conductivities. The number of integrals would continuously increase as more derivatives or generalized derivatives are applied in the perturbation theory and more ways exist to arrange them in the numerators of the $\Pi_j^{\beta\alpha_1\dots\alpha_n}$ integrals.

\section{Relaxation-free limit} \label{Relaxationfree}

In Section~\ref{Complexfrequencies}, complex frequencies were introduced as a means to allow the two-band system to relax. In this section, the limit of vanishing relaxation and real frequencies is considered.

As mentioned before, there are some subtleties in implementing this limit. For certain regions of frequency space, the nonlinear conductivity will \textit{diverge} in the relaxation-free limit. Specifically, the nonlinear conductivity may diverge in the relaxation-free limit when a subset of the optical frequencies adds to zero. This includes the case of DC fields ($\omega_i=0$) or DC currents ($\omega_1+\dots+\omega_n=0$).

Fundamentally, this is due to the existence of regions of the FBZ where more than one of the resonance conditions is obeyed simultaneously (two or more denominators in the symmetrized version of Eq.~\ref{sigmaomegabar} are resonant at those $\mathbf{k}$-points). Geometrically, this can be visualized by representing in the FBZ the set of points where the conditions $\omega_1-\Delta\epsilon_{\mathbf{k}cv}=0$ or $\omega_1+\dots+\omega_i-\Delta\epsilon_{\mathbf{k}cv}=0$ ($i=2,...,n$) and their permutations are verified and checking for overlapping regions (for example, in Fig.~\ref{fig:resonancecontours} for there to be the possibility of a divergence two of the contours would have to cross or altogether overlap).
 
When using the equations of Section~\ref{Resonancebasedanalysis}, this difficulty is realized in the relaxation-free limit of Eq.~\ref{iphoton},

\begin{equation}
    \lim_{\gamma\to 0^+}\sigma_i^{\alpha}(\bar{\omega}_1,\cdots,\bar{\omega}_n)=\lim_{\gamma\to 0^+}\sum_{j,\,p} C_{i j}^{p}(\bar{\omega}_1,\cdots,\bar{\omega}_n)\,\Pi_j^{p(\alpha)}(\bar{\omega}_1+\cdots+\bar{\omega}_i)=\sum_{j,\,p} C_{i j}^{p}(\omega_1,\cdots,\omega_n)\lim_{\gamma\to 0^+}\Pi_j^{p(\alpha)}(\bar{\omega}_1+\cdots+\bar{\omega}_i)
    \label{limit}
\end{equation}

The validity of this passage hangs on the convergence of the $C$ coefficients in the relaxation-free limit, which in turn relies on there existing no cancelling subset of frequencies in $\{\omega_1,\dots,\omega_n\}$.

Henceforth, we will assume this to be the case. We emphasize that the validity of the equations in Section~\ref{Resonancebasedanalysis} always holds, even when in these regions of frequency space. However, the analysis of the relaxation-free limit in these situations, which include several nonlinear optical effects of interest (photovoltaic effects, DC field-induced second order response, electro-optic effects,...), albeit possible, requires greater care and will be left to future communications. Indeed, the study of the divergences of the nonlinear optical conductivity is a topic of current research interest~\cite{Cheng2019IntrabandSystems,Fregoso2018JerkEffect,Ventura2021SecondSemiconductor}.

\subsection{Real and imaginary parts}\label{Realandimaginaryparts}

From Eq.~\ref{limit}, the evaluation of the relaxation-free limit of the nonlinear conductivity has been reduced to evaluating the relaxation-free limit of the integrals $\Pi_j^{\alpha}$. This is a direct application of Eq.~\ref{sokhotskiplemelj},

\begin{equation}
    \lim_{\gamma\to 0^+}\Pi_j^{\alpha}(\bar{\omega})=\pi\left(\mathcal{H}_j^{\alpha}(\omega)-i\,\mathcal{I}_j^{\alpha}(\omega)\right)
    \label{pilimit}
\end{equation}
where

\begin{align}
    \mathcal{I}_j^{\alpha}(\omega)&\equiv\int\frac{d^d\mathbf{k}}{(2\pi)^d}\sum_{a,b}g_j^{\alpha}(\mathcal{A},\epsilon)_{ab}\,\Delta f_{ba}\,\delta\left(\hbar\omega-\Delta\epsilon_{ab}\right)\label{I}\\
    \mathcal{H}_j^{\alpha}(\omega)&\equiv\dashint\frac{d^d\mathbf{k}}{(2\pi)^d}\sum_{a,b}\frac{g_j^{\alpha}(\mathcal{A},\epsilon)_{ab}}{\hbar\omega-\Delta\epsilon_{ab}}\,\Delta f_{ba}\label{H}
\end{align}

As an example, let us consider the linear conductivity. Ignoring Fermi surface contributions,

\begin{equation}
    \lim_{\gamma\to 0^+}\sigma^{\beta\alpha_1}(\bar{\omega}_1)=i e^2\,\omega_1\,\left(\mathcal{H}_1^{\beta\alpha_1}(\omega_1)-i\,\mathcal{I}_1^{\beta\alpha_1}(\omega_1)\right)
\end{equation}
where

\begin{align}
    \mathcal{I}_1^{\beta\alpha_1}(\omega_1)&=\int\frac{d^d\mathbf{k}}{(2\pi)^d}\sum_{a,b}\,\Re\{\mathcal{A}_{ba}^{\beta}\,\mathcal{A}_{ab}^{\alpha_1}\}\,\Delta f_{ba}\,\delta\left(\hbar\omega_1-\Delta\epsilon_{ab}\right)\label{I1first}\\
    \mathcal{H}_1^{\beta\alpha_1}(\omega_1)&=\dashint\frac{d^d\mathbf{k}}{(2\pi)^d}\sum_{a,b}\frac{\Re\{\mathcal{A}_{ba}^{\beta}\,\mathcal{A}_{ab}^{\alpha_1}\}}{\hbar\omega_1-\Delta\epsilon_{ab}}\,\Delta f_{ba}\label{H1first}
\end{align}

The integral $\mathcal{H}_1$ can be computed through Eq.~\ref{H1first} or from applying a Hilbert transform to $\mathcal{I}_1$. This is true in general, as it is easily proven from Eqs.~\ref{I} and~\ref{H} that

\begin{equation}
    \mathcal{H}_j^{\alpha}(\omega)=-\frac{1}{\pi}\dashint_{-\infty}^{+\infty}\frac{\mathcal{I}_j^{\alpha}(\omega')}{\omega'-\omega}\,d\omega'
\end{equation}

The entire linear response of an insulator or a cold semiconductor is obtained, for negligible relaxation, from a calculation of the function in Eq.~\ref{I1first} and its Hilbert transform. This is by no means new information in linear optics: the integral in Eq.~\ref{I1first} is nothing more than Fermi's golden rule and the Hilbert transform from $\mathcal{I}_1$ to $\mathcal{H}_1$ equates to the usual derivation of refraction from absorption by the Kramers-Kr\"{o}nig relations. Here, however, this method has been generalized to any order.

The real part of the nonlinear conductivity is given by the $\mathcal{I}$ integrals and the imaginary part by the $\mathcal{H}$ integrals\footnote{Once again, excluding the regions of frequency space where subsets of frequencies add to zero.}. This can be seen by noting that the coefficients $C$ that multiply the integrals in Eq.~\ref{limit} are purely imaginary in the relaxation-free limit, while $\mathcal{I}(\omega)$ and $\mathcal{H}(\omega)$ are always real.

The association of the real part of the nonlinear conductivity with the $\mathcal{I}$ integrals is to be expected if one recalls that the real part is the one responsible for optical absorption. If the optical frequencies (and their harmonics) of the light incident in a cold semiconductor all lie below the gap, there must be no absorption and therefore the real part of the nonlinear conductivity is required to vanish. This is guaranteed to be the case for the integrals in Eq.~\ref{I}.

We conclude from the analysis of this section that, if we wish to evaluate the real part of a nonlinear conductivity, we have only to evaluate simple integrals with the form of Eq.~\ref{I} for the appropriate $g$ functions, listed in Eqs.~\ref{FAfirst}-\ref{Pi1first},~\ref{FBsecond}-\ref{Pi2second},~\ref{FAthird}-\ref{Pi6third}. These contributions are well localized in the FBZ and run over regions that satisfy the resonance conditions.

The imaginary part of a nonlinear conductivity follows directly by performing Hilbert transforms of the integrals computed for the real part. We note that we are not performing a Hilbert transform of the entire nonlinear conductivity as it is traditionally done in the nonlinear Kramers-Kr\"{o}nig relations~\cite{Hutchings1992Kramers-KronigOptics}. We are not even applying a Hilbert transform to the different contributions in Eq.~\ref{resonancebasedanalysis} to move from their real to imaginary parts\footnote{This would not work as the real and imaginary parts of each individual contribution are not simply related by a Hilbert transform.}. The passage from the real to imaginary parts of the nonlinear optical conductivity is more subtle and is made directly through the integrals in Eq.~\ref{pilimit}. In this way, the imaginary part, which unlike the real part is not described by any restricted region of the FBZ, is made more accessible than what would perhaps be expected.

Finally, we point out that for Fermi surface contributions there is no need for the relaxation-free limit, since the $F_X$ integrals are frequency independent. To attempt an analytical calculation it is necessary only to set the temperature to zero, in which case

\begin{equation}
    F_X^{\beta\alpha_1\cdots\alpha_n}=-\int\frac{d^d\mathbf{k}}{(2\pi)^d}\sum_{a}g_X^{\alpha_1\dots\alpha_n}(\mathcal{A},\epsilon)_{aa}\,\partial^{\beta}\epsilon_a\,\delta(\mu-\epsilon_a)
    \label{fintegrals}
\end{equation}

This integral has a similar structure to the integrals in Eq.~\ref{I}, with the chemical potential taking the role of a frequency.

For the previous integrals (Eq.~\ref{I}), it is equally helpful to consider the system to be at zero temperature to make calculations tractable. However, once the nonlinear conductivity is calculated for $T=0$, it is possible to obtain the answers at finite temperature with a few tricks~\cite{Cheng2015ThirdTemperature}.

\subsection{Finite temperature} \label{Finitetemperature}

It is possible to quickly arrive at finite temperature results from zero temperature calculations of the nonlinear conductivity by making use of a relation first introduced in~\cite{Cheng2015ThirdTemperature}. The key idea is to recognize that the entire chemical potential dependence of the $\Pi_i^{\alpha}(\bar{\omega})$ and $F_X^{\alpha}$ integrals stems from the Fermi-Dirac distribution. Starting with the former type of integrals, we make all arguments of the distribution explicit,

\begin{equation}
    f(\epsilon,\mu,T)\equiv\frac{1}{e^{(\epsilon-\mu)/k_BT}+1}
    \label{fermidirac}
\end{equation}

At zero temperature,

\begin{equation}
    f(\epsilon,\mu,0)=\Theta(\mu-\epsilon)
    \label{fermidiraczero}
\end{equation}

The finite temperature and zero temperature distributions are related by

\begin{equation}
    f(\epsilon,\mu,T)=\int_{-\infty}^{+\infty}f(\mu',\mu,T)\,\delta(\mu'-\epsilon)d\mu'=\int_{-\infty}^{+\infty}f(\mu',\mu,T)\,\partial_{\mu'}f(\epsilon,\mu',0)\,d\mu'
    \label{fromzerotofinitetemperature1}
\end{equation}

Since no other objects in $\Pi_i^{\alpha}(\bar{\omega})$ depend on the chemical potential, Eq.~\ref{fromzerotofinitetemperature1} translates directly to

\begin{equation}
    \sigma_i^{\beta\alpha_1\dots\alpha_n}(\bar{\omega}_1,\dots,\bar{\omega}_n;\mu,T)=\int_{-\infty}^{+\infty}f(\mu',\mu,T)\,\,\,\partial_{\mu'}\sigma_i^{\beta\alpha_1\dots\alpha_n}(\bar{\omega}_1,\dots,\bar{\omega}_n;\mu',0)\,d\mu'
    \label{iphotonfinitetemperature}
\end{equation}
where we made the chemical potential and temperature dependencies explicit.

By a similar argument, $\sigma_F$ is proven to also obey Eq.~\ref{iphotonfinitetemperature}. Joining all contributions, we arrive at a general relation between the zero and finite temperature nonlinear conductivity, 

\begin{align}
    \sigma^{\beta\alpha_1\dots\alpha_n}(\bar{\omega}_1,\dots,\bar{\omega}_n;\mu,T)&=\int_{-\infty}^{+\infty}f(\mu',\mu,T)\,\,\,\partial_{\mu'}\sigma^{\beta\alpha_1\dots\alpha_n}(\bar{\omega}_1,\dots,\bar{\omega}_n;\mu',0)\,d\mu'\nonumber\\
    &=\frac{1}{T}\,\int_{-\infty}^{+\infty}\,(1-f(\mu',\mu,T))f(\mu',\mu,T)\,\sigma^{\beta\alpha_1\dots\alpha_n}(\bar{\omega}_1,\dots,\bar{\omega}_n;\mu',0)\,d\mu'
    \label{fromzerotofinitetemperature}
\end{align}
by partial integration, using $\sigma(\mu\rightarrow-\infty)=0$.

The effect of having a finite temperature is to probe the nonlinear conductivity at different values of the chemical potential. Eq.~\ref{fromzerotofinitetemperature} is a kind of weighted average given by a distribution that is centered and peaked at the Fermi level and has a width of the order $k_BT$.\newline

We conclude by making some observations on systems with electron-hole symmetry for which the chemical potential dependence of $\sigma_i^{\alpha}$ can be specially simple.

The real part of the one-photon contribution is given by the $\mathcal{I}_j^{\alpha}(\omega_1)$ integrals. At $T=0$ and assuming electron-hole symmetry, their integrand is proportional to,

\begin{equation}
   \Delta f_{vc}=\Theta(\mu-\epsilon_v)-\Theta(\mu-\epsilon_c)=\Theta(\Delta\epsilon_{cv}-2|\mu|)\rightarrow\Theta(\hbar|\omega_1|-2|\mu|)
   \label{deltaf}
\end{equation}
where the last step uses the Dirac delta in Eq.~\ref{I}. The final Heaviside step function is independent of \textbf{k} and can be pulled out of the $\mathcal{I}_j^{\alpha}(\omega_1)$ integrals: we have therefore an universal dependence on chemical potential, with frequency dependent coefficients set by the material, the inverse situation to what happens in $\sigma_F$. For gapped systems, we could more generally write $\Theta(\hbar|\omega_1|-\text{Max}(2|\mu|,\Delta))$, where $\text{Max}(2|\mu|,\Delta)$ is the ``effective gap''.

Similarly, for the $i$-photon contribution,

\begin{equation}
   \Delta f_{vc}\rightarrow\Theta(\hbar|\omega_1+\dots+\omega_i|-\text{Max}(2|\mu|,\Delta))
   \label{deltafiphoton}
\end{equation}

This is exemplified by the expression in Eq.~\ref{diracTHG}.

\textit{The real part of the nonlinear conductivity of electron-hole symmetric systems is always given in terms of step functions in the chemical potential}, aside from the low frequency behaviour where Fermi surface contributions dominate. These step-like variations are due to Pauli blocking. Depending on the dispersion of the coefficients of the step functions, the steps can be found to vary in magnitude and even be absent or infinite. 

On the other hand, the imaginary part is obtained by applying a Hilbert transform to $\mathcal{I}_i^{\alpha}$ and different systems will give different functions of the chemical potential. No universal behaviour can be predicted, in this context, for the imaginary part.

The corresponding finite temperature result follows from Eqs.~\ref{deltaf} and~\ref{iphotonfinitetemperature},

\begin{equation}
    \int_{-\infty}^{+\infty}f(\mu',\mu,T)\,\partial_{\mu'}\Theta(\hbar|\omega_1|-2|\mu|)\,d\mu'=f(-\hbar|\omega_1|/2,\mu,T)-f(\hbar|\omega_1|/2,\mu,T)
    \label{deltaffinitetemperature}
\end{equation}

Introducing a finite temperature smooths the step functions observed when varying the chemical potential.

\section{A simplistic model for direct gap semiconductors} \label{Parabolic}

A detailed framework for computations of the nonlinear conductivity has been described and it is now left to demonstrate it with some simple models. We consider first the case of direct gap semiconductors (3D). Of course, these materials can be not only of great interest, but of great complexity. If we focus on frequencies whose resonances lie close to the gap, however, the band structure may be approximated by a single pair of isotropic parabolic bands with effective masses,

\begin{align}
    \epsilon_{\mathbf{k}c}&=\frac{\hbar^2\,k^2}{2\,m_c}+\frac{\Delta}{2} \label{parabolicbandsc}\\
    \epsilon_{\mathbf{k}v}&=-\frac{\hbar^2\,k^2}{2\,m_v}-\frac{\Delta}{2} \label{parabolicbandsv}
\end{align}
with $k=\sqrt{k_x^2+k_y^2+k_z^2}$.

The trouble now lies in describing the non-abelian Berry connection, whose derivation relies on a knowledge of the energy eigenstates. A common approach that was adopted in early attempts at describing the nonlinear properties of solids~\cite{Butcher1991ElementsOptics,Sheik-Bahae1990DispersionAbsorption,Sheik-Bahae1991DispersionSolids,Aversa1994ThirdModel}, and is still in use in the literature~\cite{Hannes2019Higher-orderModel}, is to consider nearly constant dipole matrix elements. In the language of this work, this means assuming not only that the bands are parabolic near the gap but that the gauge field $\mathcal{A}^{\beta}$ is constant over the region of the FBZ that is being probed by the optical fields. This overly simplistic description of semiconductors has at least the merit of providing quick derivations and a first intuition on the dispersion of the nonlinear conductivity near the gap of a semiconductor.

Somewhat paradoxically, $\mathcal{A}^{\beta}$ is sometimes taken to be isotropic as well~\cite{Sheik-Bahae1990DispersionAbsorption,Sheik-Bahae1991DispersionSolids,Aversa1994ThirdModel}. This is clearly not possible since for a vector field to be isotropic and constant it either vanishes or it is not uniquely defined over the region of $\mathbf{k}$-space that is of interest. Since no degeneracies are being considered, the non-abelian Berry connection $\mathbf{\mathcal{A}}^{\beta}$ must be well-defined. We must then break the isotropy and consider a particular direction that the field is aligned with, say along the $x$ axis,

\begin{equation}
    \mathcal{A}^{\beta}_{cv}=\mathcal{A}\,\delta^{\beta x}
    \label{constantmatrixelements}
\end{equation}

Eqs.~\ref{parabolicbandsc},~\ref{parabolicbandsv} and~\ref{constantmatrixelements} define the model for which we shall compute optical conductivities. For simplicity, we consider the incident optical field to be linearly polarized along $x$ and derive the response along the same direction.

For the undoped system, the linear conductivity follows from the evaluation of the integral in Eq.~\ref{I1first},

\begin{equation}
    \mathcal{I}_1^{xx}(\omega)=|\mathcal{A}|^2\,\text{Sign}(\omega)\,\rho_{JDOS}(\hbar|\omega|)
\end{equation}
where $JDOS$ stands for joint density of states,

\begin{equation}
    \rho_{JDOS}(\hbar|\omega|)=\int\frac{d^3\mathbf{k}}{(2\pi)^3}\,\delta(\hbar|\omega|-\Delta\epsilon_{\mathbf{k}cv})=\frac{1}{\sqrt{2}\pi^2\hbar^3}\,m_r^{3/2}\,\sqrt{\hbar|\omega|-\Delta}\,\Theta\left(\hbar|\omega|-\Delta\right)
\end{equation}
and $m_r^{-1}\equiv m_c^{-1}+m_v^{-1}$.

From Eq.~\ref{sigma11},

\begin{equation}
    \Re\left(\sigma^{xx}(\omega)\right)=\frac{\pi\,e^2}{\hbar}\,\hbar\omega\,\mathcal{I}_1^{xx}(\omega)=\frac{e^2}{\sqrt{2}\pi\hbar^4}\,m_r^{3/2}\,|\mathcal{A}|^2\,\hbar|\omega|\,\sqrt{\hbar|\omega|-\Delta}\,\,\Theta\left(\hbar|\omega|-\Delta\right)
\end{equation}

\begin{figure}
    \centering
    \begin{subfigure}{0.49\textwidth}
        \centering
        \includegraphics[width=\linewidth]{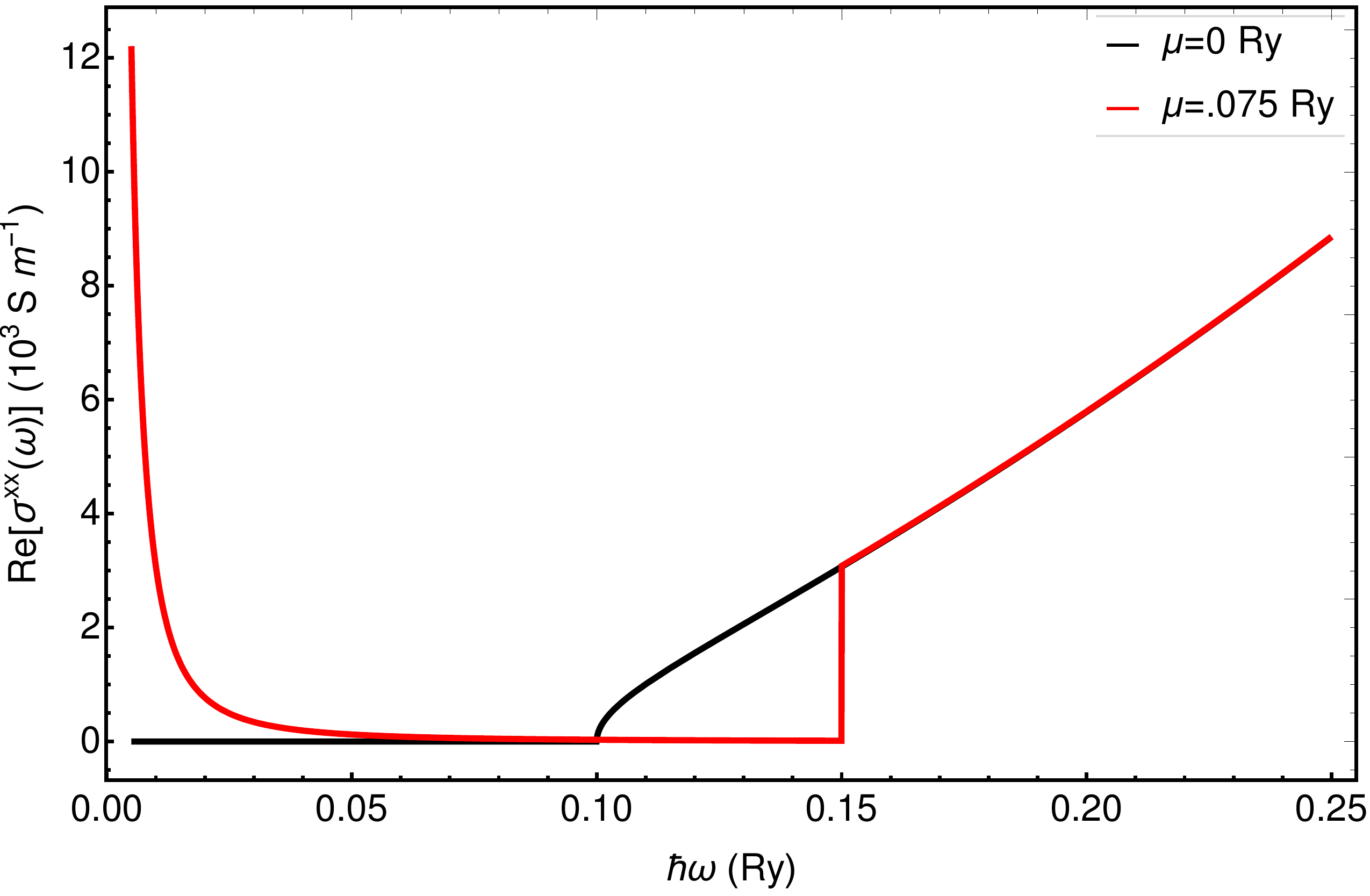}
        \caption{}
        \label{fig:parabolicbands1}
    \end{subfigure}
    \,
    \begin{subfigure}{0.49\textwidth}
        \centering
        \includegraphics[width=\linewidth]{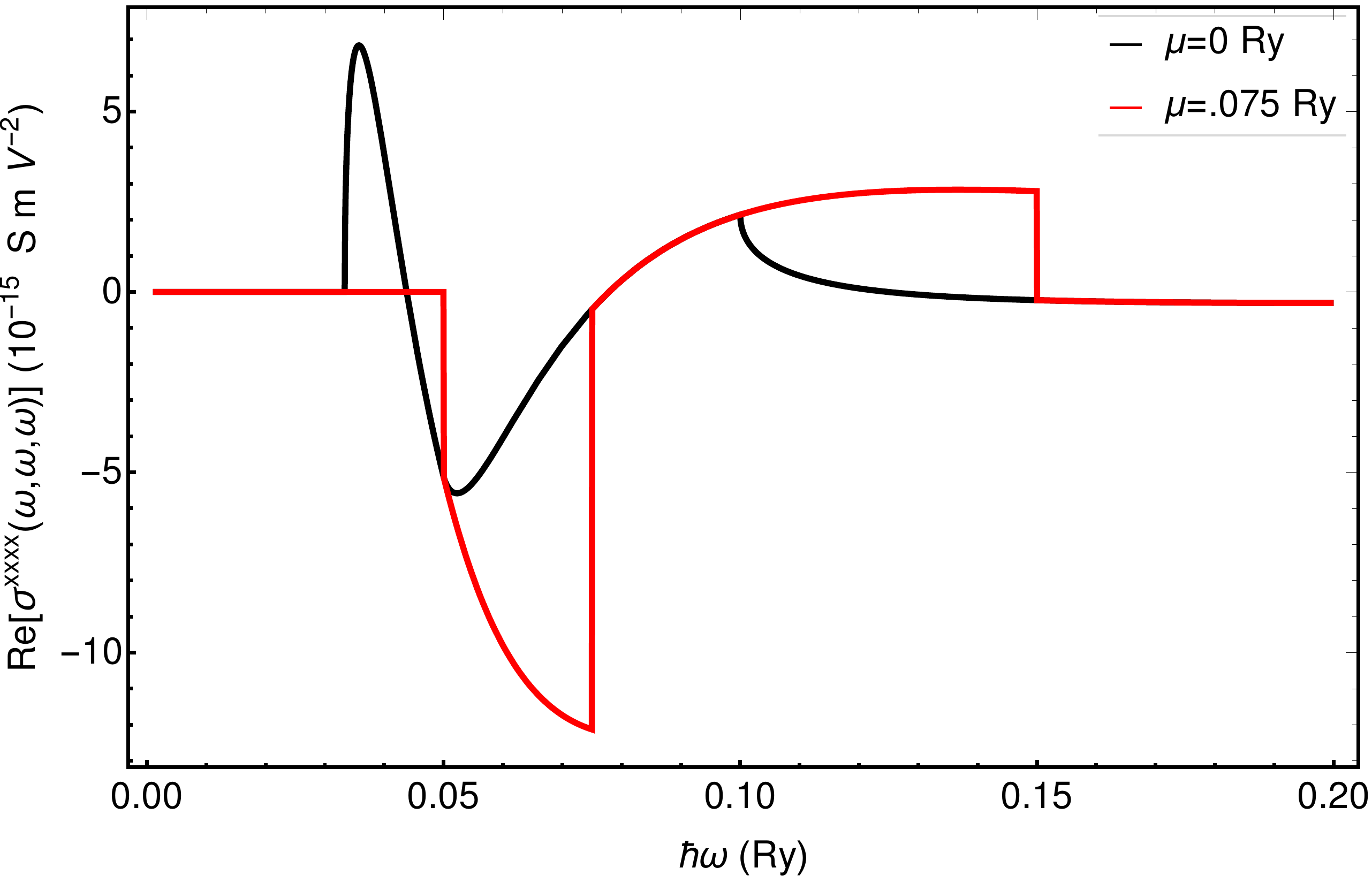}
        \caption{}
        \label{fig:parabolicbands2}
    \end{subfigure}
    \caption{\small{Real part of the (a) linear and (b) third order conductivity for a system of parabolic bands. The energies are given in Rydberg's ($1\,\text{Ry}\simeq 13.6\,eV$). The chosen parameters are: $m_c=m_v=0.1\,m_e$, $\mathcal{A}=6.5\,\text{\AA}$ and $\Delta=0.1\,\text{Ry}$. Both the undoped (black) and doped (red) curves are represented, with the effective gap set at $0.15\,\text{Ry}$ in the doped case. The value chosen for $\mathcal{A}$ is derived from the value of $E_p$ in~\cite{Sheik-Bahae1991DispersionSolids}.}}
    \label{fig:parabolic}
\end{figure}

The real part of the linear conductivity is depicted in Fig.~\ref{fig:parabolicbands1}. The dispersion observed near the gap is directly related to the well-known square-root dependence of optical absorption on frequency~\cite{Butcher1991ElementsOptics}.

Since the joint density of states vanishes at the gap, there are no steps in the dispersion of the linear conductivity until the system is doped and an effective gap is set at higher energies (Fig.~\ref{fig:parabolicbands1}). The linear conductivity is then given by\footnote{The Fermi surface contribution gives the Drude conductivity, whose real part consists of a Dirac delta at zero frequency in the relaxation-free limit. Since we excluded zero frequencies from our analysis (see the discussion in Section~\ref{Relaxationfree}) the real part of the Fermi surface contribution is neglected in Eqs.~\ref{sigma1parabolic},~\ref{sigmaFlineargraphene} and~\ref{sigmaFgraphene}.}

\begin{equation}
    \Re\left(\sigma^{xx}(\omega)\right)=\Re\left(\sigma_1^{xx}(\omega)\right)=\frac{e^2}{\sqrt{2}\pi\hbar^4}\,m_r^{3/2}\,|\mathcal{A}|^2\,\hbar|\omega|\,\sqrt{\hbar|\omega|-\Delta}\,\,\Theta\left(\hbar|\omega|-\Delta_{eff}\right)
    \label{sigma1parabolic}
\end{equation}
with the effective gap,

\begin{equation}
    \Delta_{eff}=
    \begin{cases}
    \frac{m_c}{m_r}\left(\mu-\Delta/2\right)+\Delta & \mu>0 \\
    \frac{m_v}{m_r}\left(-\mu-\Delta/2\right)+\Delta & \mu\leq 0
    \end{cases}
    \label{effectivegap}
\end{equation}

To obtain the imaginary part, we would normally proceed via the Kramers-Kr\"{o}nig relations, but we are faced with a challenge that is inherent to low energy descriptions of solids: the Hilbert transforms do not necessarily converge. This is the case here, as evaluation of Eq.~\ref{H1first} gives infinity. A proper model defined over the entire FBZ would always be bounded in energy and the Hilbert transforms are then guaranteed to converge. But for this very basic model, the imaginary part is inaccessible.

Similar remarks can be made for the third order conductivity (the second order conductivity vanishes due to the properties of inversion symmetry respected by Eqs.~\ref{parabolicbandsc},~\ref{parabolicbandsv} and~\ref{constantmatrixelements}). Since the $\mathbf{k}$ dependence of $\mathcal{A}$ is neglected, the generalized derivatives are taken to be identically zero and half of the integrals to be evaluated at third order vanish,

\begin{equation}
    \mathcal{I}_1^{xxxx}(\omega)=\mathcal{I}_4^{xxxx}(\omega)=\mathcal{I}_6^{xxxx}(\omega)=0
\end{equation}

The remaining integrals are straightforwardly computed (Eqs.~\ref{Pi2third},~\ref{Pi3third} and~\ref{Pi5third}),

\begin{align}
    \mathcal{I}_2^{xxxx}(\omega)&=\frac{\hbar^2}{m_r}\,|\mathcal{A}|^2\,\rho_{JDOS}(\hbar|\omega|)=\frac{1}{\sqrt{2}\pi^2\hbar}\,m_r^{1/2}\,|\mathcal{A}|^2\,\sqrt{\hbar|\omega|-\Delta}\,\Theta\left(\hbar|\omega|-\Delta_{eff}\right)\label{I2parabolic}\\
    \mathcal{I}_3^{xxxx}(\omega)&=\frac{\sqrt{2}}{3\pi^2\hbar}\,m_r^{1/2}\,|\mathcal{A}|^2\,\text{Sign}(\omega)\,\left(\hbar|\omega|-\Delta\right)^{3/2}\,\,\Theta\left(\hbar|\omega|-\Delta_{eff}\right)\label{I3parabolic}\\
    \mathcal{I}_5^{xxxx}(\omega)&=|\mathcal{A}|^4\,\text{Sign}(\omega)\,\rho_{JDOS}(\hbar|\omega|)=\frac{1}{\sqrt{2}\pi^2\hbar^3}\,m_r^{3/2}\,|\mathcal{A}|^4\,\text{Sign}(\omega)\,\sqrt{\hbar|\omega|-\Delta}\,\Theta\left(\hbar|\omega|-\Delta_{eff}\right)\label{I5parabolic}
\end{align}

Direct substitution of these integrals in the relaxation-free limit of Eqs.~\ref{sigma3}-\ref{sigma33} immediately gives a general expression for the third order conductivity. For brevity, we make the expression explicit only for the special case of harmonic generation ($\omega_1=\omega_2=\omega_3=\omega$),

\begin{equation}
    \sigma^{xxxx}(\omega,\omega,\omega)=\sigma_1^{xxxx}(\omega,\omega,\omega)+\sigma_2^{xxxx}(\omega,\omega,\omega)+\sigma_3^{xxxx}(\omega,\omega,\omega)    
\end{equation}
where the one-, two- and three-photon contributions are

\begin{align}
    \sigma_1^{xxxx}(\omega,\omega,\omega)&=\frac{\pi i e^4}{\hbar}\left(-\frac{5\,\mathcal{I}_2^{xxxx}(\omega)}{4\,(\hbar\omega)^2}-\frac{11\,\mathcal{I}_3^{xxxx}(\omega)}{6\,(\hbar\omega)^3}+\frac{3\,\mathcal{I}_5^{xxxx}(\omega)}{4\,\hbar\omega}\right)\label{parabolicsigma1}\\
    \sigma_2^{xxxx}(\omega,\omega,\omega)&=\frac{\pi i e^4}{\hbar}\,\frac{10\,\mathcal{I}_3^{xxxx}(2\omega)}{3\,(\hbar\omega)^3}\label{parabolicsigma2}\\
    \sigma_3^{xxxx}(\omega,\omega,\omega)&=\frac{\pi i e^4}{\hbar}\left(\frac{3\,\mathcal{I}_2^{xxxx}(3\omega)}{4\,(\hbar\omega)^2}-\frac{3\,\mathcal{I}_3^{xxxx}(3\omega)}{2\,(\hbar\omega)^3}-\frac{3\,\mathcal{I}_5^{xxxx}(3\omega)}{4\,\hbar\omega}\right)\label{parabolicsigma3}
\end{align}

The Fermi surface contributions are absent at third order for a quadratic dispersion relation ($\partial^3\epsilon=0$ in Eq.~\ref{FAthird}).

The real part of the third order conductivity is represented in Fig.~\ref{fig:parabolicbands2}. Three features are clearly seen for increasing frequency when the three-, two- and one-photon contributions, respectively, become relevant. For the doped case, the usual steps are observed, stemming from the Heaviside theta functions in Eqs.~\ref{I2parabolic},~\ref{I3parabolic} and~\ref{I5parabolic}, which in turn are manifestations of Pauli blocking of transitions below the Fermi level. In the undoped case, analogous to linear order, the steps are suppressed by the vanishing joint density of states, giving rise to pronounced, but less abrupt features.

Once again, the imaginary part is not captured by the current description.

\section{Nonlinear conductivity of graphene near the Dirac point} \label{Monolayergraphene}

\begin{figure}[t]
    \centering
    \begin{subfigure}{0.45\textwidth}
        \centering
    \includegraphics[width=0.5\linewidth]{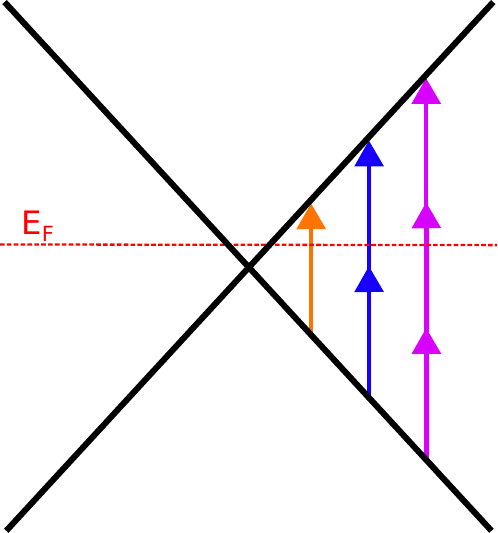}
        \caption{}
        \label{fig:diraccone}
    \end{subfigure}
        \begin{subfigure}{0.45\textwidth}
        \centering
        \includegraphics[width=0.4\linewidth]{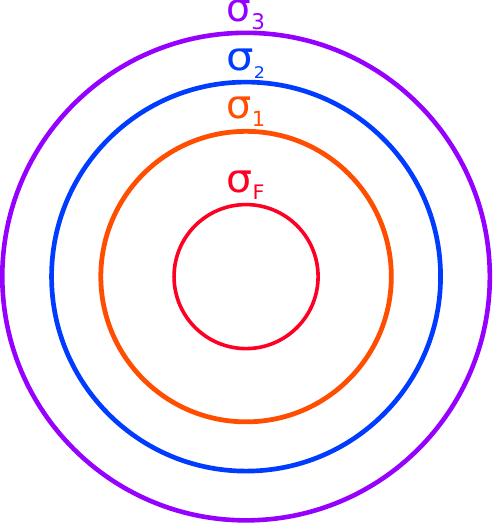}
        \caption{}
        \label{fig:resonancecontours}
    \end{subfigure}
    \caption{\small{(a) Cross section of the Dirac cone~\cite{CastroNeto2009TheGraphene} with the dark lines showing the linear energy-momentum relation in Eqs.~\ref{conductionbandgraphene} and~\ref{valencebandgraphene}. The arrows stand for incident photons, all of which are assumed to have the same energy (the color scheme serves only to differentiate the different contributions) but probe different regions of the FBZ depending on the number of photons involved. (b) Contours in the FBZ where the resonance conditions that define the different contributions in Eq.~\ref{resonancebasedanalysis} are met. For the two-dimensional crystal of monolayer graphene, they consist of circles centered at the Dirac point. The red circle is the Fermi surface.}}
\end{figure}

As a final example, we turn to monolayer graphene~\cite{Katsnelson2007Graphene:Dimensions,CastroNeto2009TheGraphene} whose nonlinear optical properties have been a subject of active research since the pioneering works of Mikhailov et al.~\cite{Mikhailov2007Non-linearGraphene,Mikhailov2008NonlinearEffects,Hendry2010CoherentGraphene}. A widely adopted two-band tight binding model provides the band structure and non-abelian Berry connection near the Dirac point~\cite{Cheng2014ThirdGraphene},

\begin{align}
    \epsilon_{\mathbf{k}c}&=+v_F\,\hbar\,k\label{conductionbandgraphene}\\
    \epsilon_{\mathbf{k}v}&=-v_F\,\hbar\,k\label{valencebandgraphene}\\
    \mathcal{A}^{\beta}_{cv}&=\frac{(\mathbf{\hat{z}}\cross\mathbf{k})^{\beta}}{2\,k}\label{nonabelianberryconnection}
\end{align}
with $\mathbf{\hat{z}}=(0,0,1)$, $k=\sqrt{k_x^2+k_y^2}$ and $v_F\simeq 10^6\,m/s$ as the Fermi velocity. The energy-momentum relation is depicted in Fig.~\ref{fig:diraccone} together with the contours defined by photon resonances in Fig.~\ref{fig:resonancecontours}

We leave the integral evaluation to an appendix (Appendix \hyperlink{D}{D}) and proceed to inspect the analytic expressions for the conductivities. In linear order, symmetry reduces the number of independent tensor elements to one,

\begin{align}
    \sigma^{xy}(\omega)&=\sigma^{yx}(\omega)=0\\
    \sigma^{yy}(\omega)&=\sigma^{xx}(\omega)=\sigma_F^{xx}(\omega)+\sigma_1^{xx}(\omega)
\end{align}
with

\begin{align}
    \sigma_F^{xx}(\omega)&=\frac{4\,i\,\sigma_0}{\pi}\,\frac{\mu}{\hbar\omega}\label{sigmaFlineargraphene}\\
    \sigma_1^{xx}(\omega)&=\sigma_0\left(\Theta\left(\hbar|\omega|-2|\mu|\right)+\frac{i}{\pi}\ln\left|\frac{\hbar\omega-2|\mu|}{\hbar\omega+2|\mu|}\right|\right)\label{sigma1xx}
\end{align}

For $|\omega|\gg 2|\mu|$, we find the universal conductivity of graphene $\sigma^{xx}(\omega)=\sigma_0=e^2/4\hbar$~\cite{Nair2008FineGraphene}.

Fortunately, in this model the Hilbert transforms converge (Eqs.~\ref{Hilberttransformlinear} and~\ref{Hilberttransform} of Appendix~\hyperlink{D}{D}) and expressions for the imaginary part are obtainable, as included in Eq.~\ref{sigma1xx}. The case of graphene showcases a general characteristic of the optical response that stems from the Hilbert transforms relating $\mathcal{I}$ and $\mathcal{H}$: \textit{discontinuities in the real part} (as those introduced by Heaviside theta functions) \textit{translate into logarithmic divergences in the imaginary part} and vice-versa.

Monolayer graphene is inversion symmetric so it has no second order response.

\begin{figure}[t]
    \centering
    \begin{subfigure}{0.49\textwidth}
        \centering
        \includegraphics[width=\linewidth]{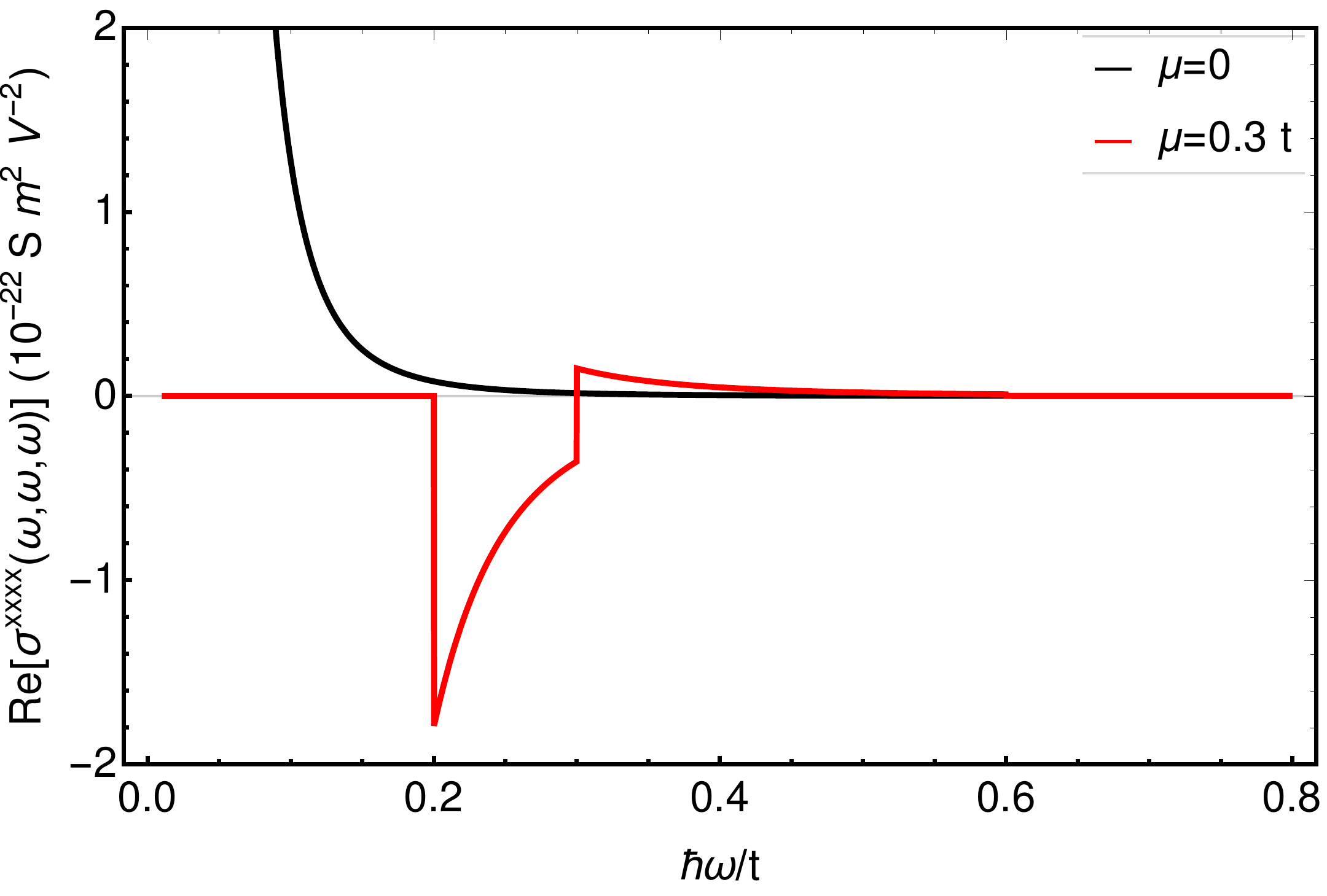}
        \caption{}
        \label{fig:THGrealpart}
    \end{subfigure}
    \,
    \begin{subfigure}{0.49\textwidth}
        \centering
        \includegraphics[width=\linewidth]{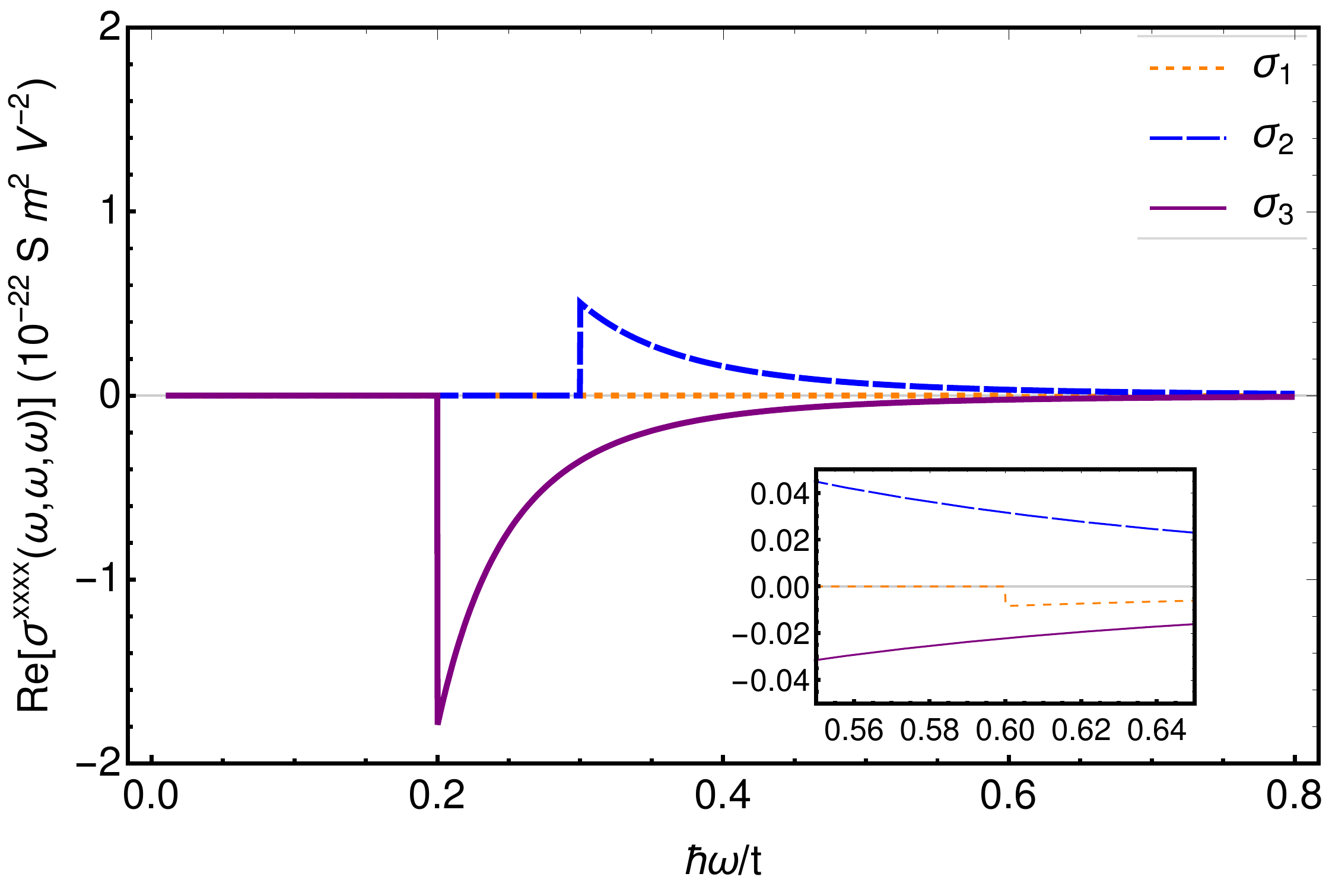}
        \caption{}
        \label{fig:THGrealpartdecomposed}
    \end{subfigure}
    \caption{\small{Real part of the third order conductivity of monolayer graphene as a function of the optical frequency, normalized to the tight binding parameter $t\simeq 3 eV$~\cite{CastroNeto2009TheGraphene}, near the Dirac point. In (a), both undoped and doped graphene are shown. In (b), the different contributions to the conductivity (see Eq.~\ref{resonancebasedanalysis} and Fig.~\ref{fig:resonancecontours}) are represented for $\mu=0.3\,t$. In the inset, it is possible to discern a small feature in the one-photon contribution.}}
    \label{fig:THGreal}
\end{figure}

Evaluation of the integrals in Eqs.~\ref{FAthird}-\ref{Pi6third} gives a third order conductivity (Appendix~\hyperlink{D}{D}) in agreement with the results in the literature~\cite{Cheng2014ThirdGraphene}. The third order response of graphene has already been the subject of several theoretical studies~\cite{Mikhailov2007Non-linearGraphene,Mikhailov2008NonlinearEffects,Cheng2014ThirdGraphene,Cheng2015ThirdTemperature,Cheng2019ThirdApproximation} and the form of the third order conductivity in the independent electron approximation is known~\cite{Cheng2014ThirdGraphene,Jiang2018Gate-tunableGraphene}. Here, we merely aim to exemplify how the equations of Section~\ref{Resonancebasedanalysis} provide a straightforward derivation and how the structure of the nonlinear conductivity emerges naturally in this framework.

For these purposes, it is sufficient to consider third harmonic generation where symmetry constraints reduce the tensor to a single independent component $\sigma^{xxyy}(\omega,\omega,\omega)=\sigma^{xyxy}(\omega,\omega,\omega)=\sigma^{xyyx}(\omega,\omega,\omega)=\sigma^{xxxx}(\omega,\omega,\omega)/3$~\cite{Cheng2014ThirdGraphene,Jiang2018Gate-tunableGraphene}.

Using the methods of Sections~\ref{Resonancebasedanalysis} and~\ref{Relaxationfree}, we find the third order conductivity has the form,

\begin{equation}
    \sigma^{xxxx}(\omega,\omega,\omega)=\sigma_F^{xxxx}(\omega,\omega,\omega)+\sigma_1^{xxxx}(\omega,\omega,\omega)+\sigma_2^{xxxx}(\omega,\omega,\omega)+\sigma_3^{xxxx}(\omega,\omega,\omega)
    \label{sigmaxxxx}
\end{equation}
with

\begin{align}
    \sigma_F^{xxxx}(\omega,\omega,\omega)&=\frac{24\,i\,\hbar\,C_0}{\pi|\mu|\omega^3}\label{sigmaFgraphene}\\
    \sigma_1^{xxxx}(\omega,\omega,\omega)&=-\frac{17\,C_0}{\omega^4}\left(\Theta\left(\hbar|\omega|-2|\mu|\right)+\frac{i}{\pi}\ln\left|\frac{\hbar\omega-2|\mu|}{\hbar\omega+2|\mu|}\right|\right)+\frac{71\,i\,\hbar\,C_0}{\pi|\mu|\omega^3}\label{sigma1graphene}\\
    \sigma_2^{xxxx}(\omega,\omega,\omega)&=+\frac{64\,C_0}{\omega^4}\left(\Theta\left(2\hbar|\omega|-2|\mu|\right)+\frac{i}{\pi}\ln\left|\frac{2\hbar\omega-2|\mu|}{2\hbar\omega+2|\mu|}\right|\right)\label{sigma2graphene}-\frac{32\,i\,\hbar\,C_0}{\pi|\mu|\omega^3}\\
    \sigma_3^{xxxx}(\omega,\omega,\omega)&=-\frac{45\,C_0}{\omega^4}\left(\Theta\left(3\hbar|\omega|-2|\mu|\right)+\frac{i}{\pi}\ln\left|\frac{3\hbar\omega-2|\mu|}{3\hbar\omega+2|\mu|}\right|\right)-\frac{63\,i\,\hbar\,C_0}{\pi|\mu|\omega^3}\label{sigma3graphene}
\end{align}

\begin{figure}[t]
    \centering
    \begin{subfigure}{0.49\textwidth}
        \centering
        \includegraphics[width=\linewidth]{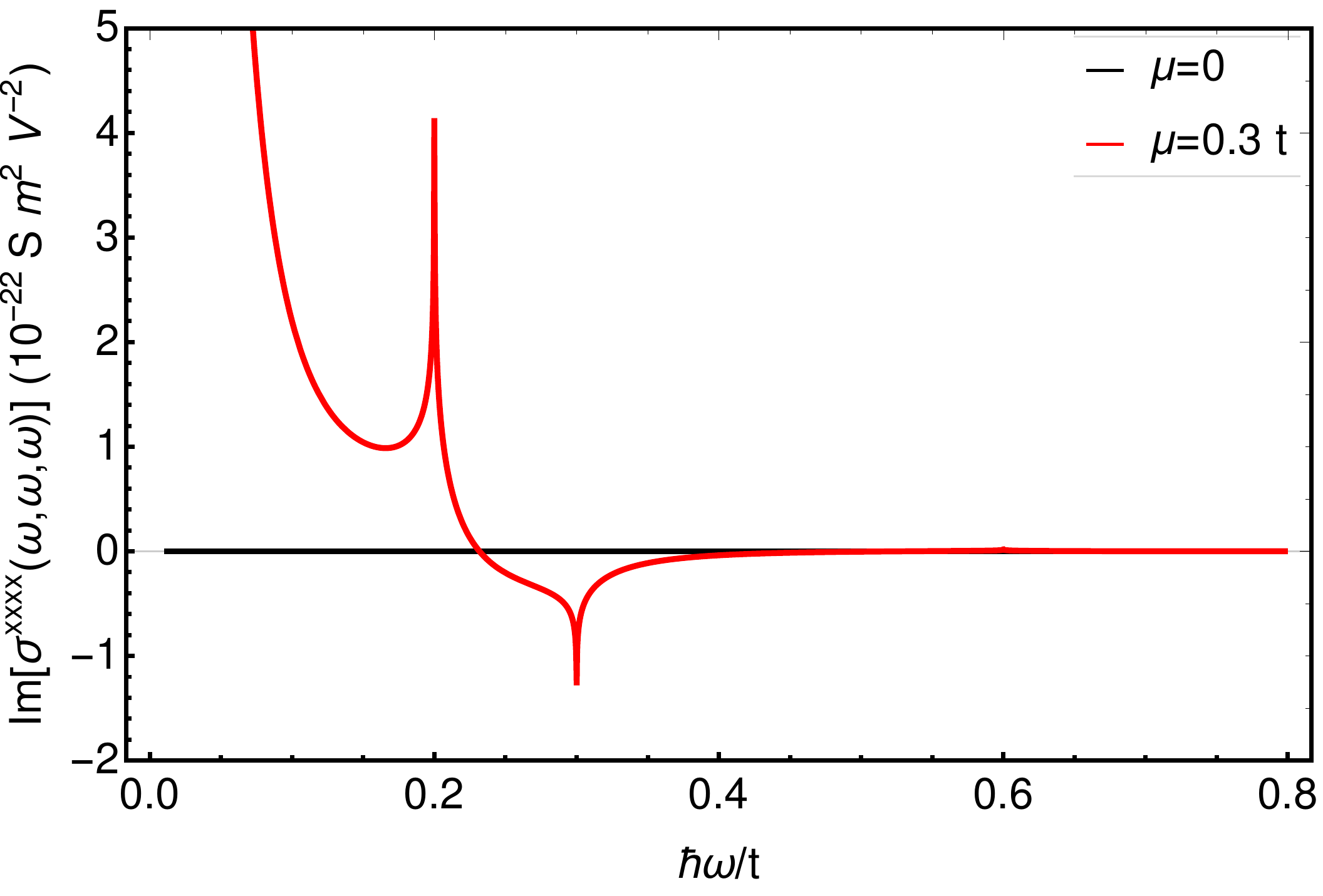}
        \caption{}
        \label{fig:THGimaginarypart}
    \end{subfigure}
    \,
    \begin{subfigure}{0.49\textwidth}
        \centering
        \includegraphics[width=\linewidth]{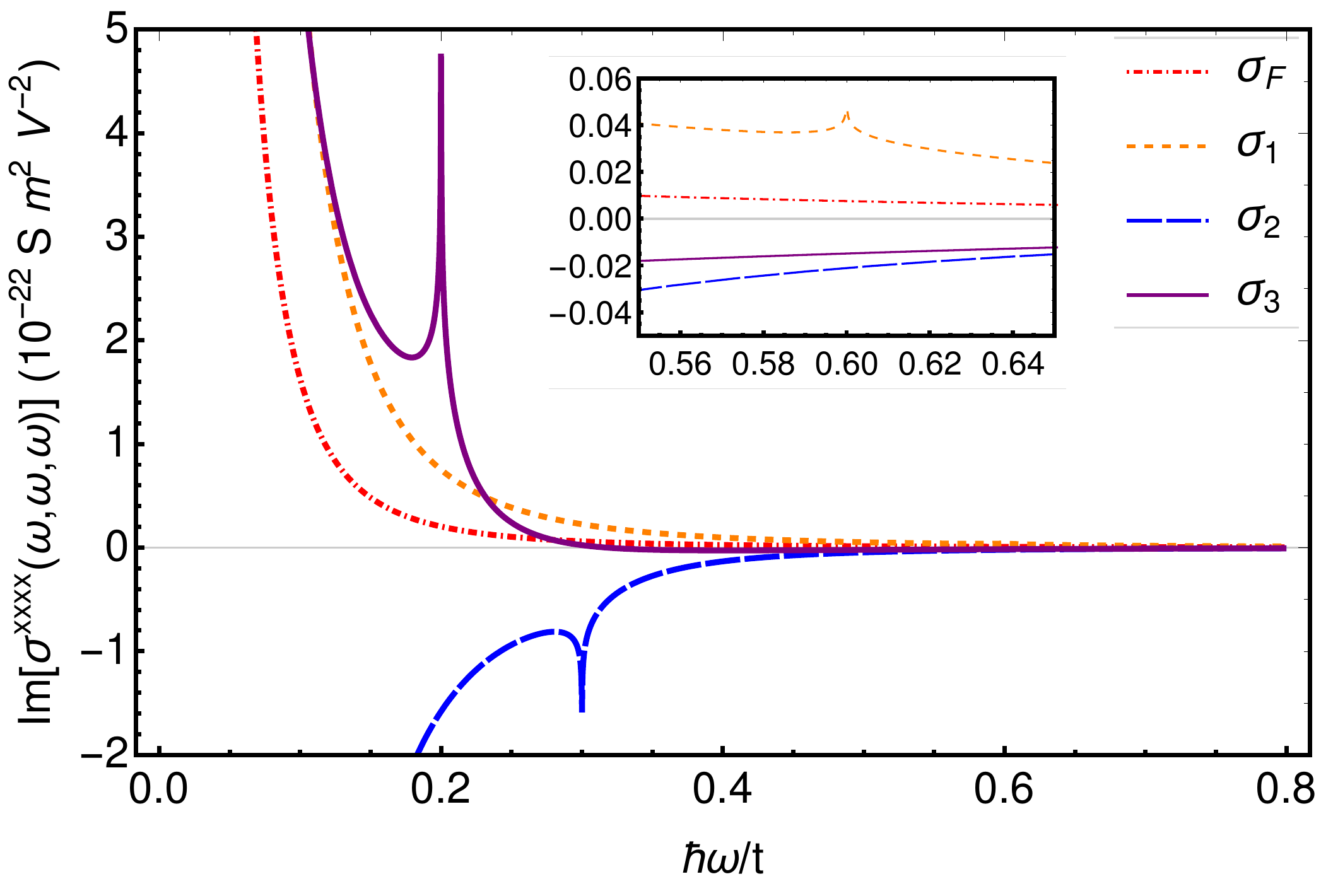}
        \caption{}
        \label{fig:THGimaginarypartdecomposed}
    \end{subfigure}
    \caption{\small{Imaginary part of the third order conductivity of monolayer graphene as a function of the optical frequency, normalized to the tight binding parameter $t\simeq 3 eV$~\cite{CastroNeto2009TheGraphene}, near the Dirac point. In (a), both undoped and doped graphene are shown. In (b), the different contributions to the conductivity (see Eq.~\ref{resonancebasedanalysis} and Fig.~\ref{fig:resonancecontours}) are represented for $\mu=0.3\,t$. In the inset, it is possible to discern a small feature in the one-photon contribution.}}
    \label{fig:THGimaginary}
\end{figure}

The result is plotted in Figs.~\ref{fig:THGreal} and~\ref{fig:THGimaginary}. Inspection of Eqs~\ref{sigma1graphene}-\ref{sigma3graphene} and the resulting curves shows that the different contributions are, as expected, dominant in the specific regions of the spectrum where the relevant photon frequencies match the gap. Once again the real part of the conductivity is marked by jump discontinuities and the imaginary part by the associated logarithmic divergences. The real parts of the one-, two- and three-photon contributions are zero below the gap, while the imaginary parts tend to zero in the DC limit. In this limit, the Fermi surface contribution (Eq.~\ref{sigmaFgraphene}) is dominant.

A curious special case is that of undoped graphene. A hasty look through the expressions for the third order conductivity might lead one to think that it diverges when the chemical potential is set to zero. But when summing all the contributions (Eqs.~\ref{sigmaFgraphene}-\ref{sigma3graphene}) to the third order conductivity the different factors of $\mu^{-1}$ cancel (Eq.~\ref{diracTHG}).

Nonetheless, there have been some concerns with this divergence in the literature. In the work of Cheng et al.~\cite{Cheng2015ThirdTemperature} a divergence of such nature was found. But, as pointed out by the authors, it appears only when introducing two distinct damping parameters, one for intraband and one for interband resonances. This suggests that such a divergence is an artifact of the phenomenological description of relaxation that was adopted in their work. In the method used here, there is a single parameter $\gamma$ describing relaxation, but this parameter might be made different at low and high frequencies (as pointed out before, any even function $\gamma=\gamma(\omega)$ will do) and still no divergences for $\mu=0$ are found.

Similarly, for the Boltzmann equation approach adopted in the pioneering works of~\cite{Mikhailov2007Non-linearGraphene,Mikhailov2008NonlinearEffects}, the third order conductivity was found to diverge when $\mu\rightarrow 0$. The reason is transparent in our framework: the Boltzmann equation approach takes only the contribution from the Fermi surface into account (purely intraband transitions) and this one can be seen from Eq.~\ref{sigmaFgraphene} to indeed diverge for a zero chemical potential. Boltzmann equation methods are appropriate to describe the low frequency nonlinear conductivity as long as there is a gap, either a real or an effective one set by Pauli blocking, below which all contributions other than the ones defined by the Fermi surface become increasingly negligible. For undoped graphene however, this is never the case and the remaining contributions in Eq.~\ref{sigmaxxxx} must be considered.

Unfortunately, a real difficulty can still be found for undoped systems with zero gap in the DC limit. Even though the $\mu=0$ third order conductivity of graphene is well defined for any finite frequency, it shows a frequency dependence $\omega^{-4}$ (see Eq.~2 of~\cite{Cheng2014ThirdGraphene}) diverging when $\omega\rightarrow 0$. A similar point could be made about the model in the previous section if we close the gap ($\Delta_{eff}=\Delta=0$ in Eqs.~\ref{I2parabolic}-\ref{I5parabolic}). This result holds even for a tight binding model defined over the entire FBZ and is presumably due to a breakdown in perturbation theory.

\section{Conclusions} \label{Conclusions}

A decomposition of the nonlinear conductivity based on the possible resonances between optical frequencies and band energies is useful, both from the perspective of practical calculations as well as in developing an physical intuition and an understanding of the structure of this response function. This was exemplified with an extensive discussion of the nonlinear optical response of the time-reversal symmetric two-band crystal.

Calculations of the first, second and third order conductivity have been shown to reduce to the evaluation of a small number of integrals over the FBZ. Evaluation of these integrals may be performed numerically but also analytically if one considers the relaxation-free limit. In this limit, obtaining the real part of a nonlinear conductivity is no harder than performing a series of Fermi golden rule computations. The imaginary part follows by applying Hilbert transforms directly to the results of these computations, without need of the nonlinear Kramers-Kr\"{o}nig relations.

A known difficulty in implementing the relaxation-free limit is that the nonlinear conductivity is then singular for extended regions of the frequency plane. These singularities are behind many of the more interesting optical effects that have been studied in recent years, such has the shift, injection and jerk currents~\cite{VanDriel1997CoherenceSemiconductors,Fregoso2018JerkEffect}. For brevity, we did not address these effects here, but the equations in Section~\ref{Resonancebasedanalysis} provide a natural starting point for the derivation of expressions for small but finite $\gamma$.

The effect of temperature was considered and some simple models solved for the purpose of illustrating the method. The third order conductivity of monolayer graphene was shown to be well-defined in the limit $\mu\rightarrow 0$ and the connection was made to the difficulties found in Boltzmann equation calculations.

We tried to emphasize how this restructuring of nonlinear conductivity allows general results to emerge more clearly: the absence of a Drude peak for the diagonal tensor elements of the second order conductivity; the universal frequency dependence that comes from Fermi surface contributions (dominant for frequencies that lie much below the gap); the step functions that determine the chemical potential dependence of the real part of the nonlinear conductivity; and so on. Further statements can be made when considering the constraints set by symmetry; these will be covered in a subsequent publication. It is our hope that the treatment presented here may facilitate future investigations of the nonlinear optical properties of solids that operate within the constraints of the independent electron and electric dipole approximations.\newline

\section*{Acknowledgments}

The authors acknowledge funding from Fundação da Ciência e Tecnologia (FCT), under the COMPETE 2020 program in FEDER component (European Union), through projects POCI-01-0145-FEDER-028887, UID/FIS/04650/2019 and M-ERA-NET2/0002/2016. The work of G. B. V. and D. J. P. is supported by the grants PD/BD/140891/2018 and PD/BD/135019/2017, respectively.

\newpage

\appendix

\section*{\hypertarget{A}{Appendix A}: Commuting position operators}

It is implicitly assumed throughout this work that position operators commute. This may seem a trivial statement, derived from the basics of quantum mechanics, but commutation relations tend to be broken upon band truncation and when a finite band model is used to describe a system, defined perhaps by the specification of a set of Bloch states $\psi_{\mathbf{k}a}$ and bands $\epsilon_{\mathbf{k}a}$, there is no reason to assume that the following will be true,

\begin{equation}
    \left[\hat{r}^{\alpha},\hat{r}^{\beta}\right]=-\left[\hat{D}^{\alpha},\hat{D}^{\beta}\right]=0
    \label{covariantcommmutator}
\end{equation}

Still, this particular commutation relation can be shown to hold on very general grounds,

\begin{align}
    \left[\hat{D}^{\alpha},\hat{D}^{\beta}\right]_{\mathbf{k}ab}&=  -i\,\partial^{\alpha}\mathcal{A}^{\beta}_{\mathbf{k}ab}+i\,\partial^{\beta}\mathcal{A}^{\alpha}_{\mathbf{k}ab}-\left[\mathcal{A}^{\alpha},\mathcal{A}^{\beta}\right]_{\mathbf{k}ab}\nonumber\\
    &=\partial^{\alpha}\left(\bra{u_{\mathbf{k}a}}\ket{\partial^{\beta}u_{\mathbf{k}b}}\right)-\partial^{\beta}\left(\bra{u_{\mathbf{k}a}}\ket{\partial^{\alpha}u_{\mathbf{k}b}}\right)+\sum_c\left(\bra{u_{\mathbf{k}a}}\ket{\partial^{\alpha}u_{\mathbf{k}c}}\bra{u_{\mathbf{k}c}}\ket{\partial^{\beta}u_{\mathbf{k}b}}-\bra{u_{\mathbf{k}a}}\ket{\partial^{\beta}u_{\mathbf{k}c}}\bra{u_{\mathbf{k}c}}\ket{\partial^{\beta}u_{\mathbf{k}b}}\right)\nonumber\\
    &=\bra{\partial^{\alpha}u_{\mathbf{k}a}}\ket{\partial^{\beta}u_{\mathbf{k}b}}-\bra{\partial^{\beta}u_{\mathbf{k}a}}\ket{\partial^{\alpha}u_{\mathbf{k}b}}-\sum_c\left(\bra{\partial^{\alpha}u_{\mathbf{k}a}}\ket{u_{\mathbf{k}c}}\bra{u_{\mathbf{k}c}}\ket{\partial^{\beta}u_{\mathbf{k}b}}-\bra{\partial^{\beta}u_{\mathbf{k}a}}\ket{u_{\mathbf{k}c}}\bra{u_{\mathbf{k}c}}\ket{\partial^{\alpha}u_{\mathbf{k}b}}\right)\nonumber\\
    &=0
\end{align}
with the last step making use of the closure relation,

\begin{equation}
    \sum_c\ket{u_{\mathbf{k}c}}\bra{u_{\mathbf{k}c}}=\hat{\mathbbm{1}}
    \label{closure}
\end{equation}

Eq.~\ref{closure} is the essential assumption in the construction of the finite-band model that ensures the validity of Eq.~\ref{covariantcommmutator}. With the commutation of position operators on firm grounds, some of its consequences can be discussed.

A curious and sometimes useful way to rewrite the statement of Eq.~\ref{covariantcommmutator} is

\begin{equation}
    \left[\hat{D}^{\alpha},\hat{\mathcal{A}}^{\beta}\right]_{ab}=\partial^{\beta}\mathcal{A}^{\alpha}_{ab}
    \label{BerryDerivative}    
\end{equation}
where we dropped the $\mathbf{k}$ label.

From Eq.~\ref{covariantcommmutator}, various other identities can be derived. Expanding the commutator,

\begin{equation}
    \partial^{\alpha}\mathcal{A}^{\beta}_{ab}-\partial^{\beta}\mathcal{A}^{\alpha}_{ab}=i\,\left[\mathcal{A}^{\alpha},\mathcal{A}^{\beta}\right]_{ab}
    \label{firstidentity}
\end{equation}

In the particular case that $a=b$, it relates the abelian Berry curvature to the off-diagonal matrix elements of the Berry connection,

\begin{equation}
    \mathcal{F}^{\alpha\beta}\equiv\partial^{\alpha}\mathcal{A}^{\beta}_{aa}-\partial^{\beta}\mathcal{A}^{\alpha}_{aa}=i\,\sum_{c\neq a,b}\left(\mathcal{A}^{\alpha}_{ac}\mathcal{A}^{\beta}_{ca}-\mathcal{A}^{\beta}_{ac}\mathcal{A}^{\alpha}_{ca}\right)
    \label{berrycurvatureandcommutator}
\end{equation}

For a two-band model, Eq.~\ref{firstidentity} with $a\neq b$ relates ``generalized derivatives'' of Berry connections,

\begin{equation}
    \mathcal{A}^{\beta}_{ab;\alpha}=\mathcal{A}^{\alpha}_{ab;\beta}
    \label{identity2}
\end{equation}

A more sophisticated identity on second order ``generalized derivatives'' can also be derived,

\begin{equation}
    \mathcal{A}^{\beta}_{ab;\alpha_1\alpha_2}-\mathcal{A}^{\beta}_{ab;\alpha_2\alpha_1}=2\left(\mathcal{A}^{\alpha_1}_{ba}\mathcal{A}^{\beta}_{ab}\mathcal{A}^{\alpha_2}_{ab}-\mathcal{A}^{\alpha_2}_{ba}\mathcal{A}^{\beta}_{ab}\mathcal{A}^{\alpha_1}_{ab}\right)
\end{equation}
which translates into a relation between $\Pi_4$ and $\Pi_5$ which are no longer independent,

\begin{equation}
    \Pi_4^{\beta\alpha_1\alpha_2\alpha_3}(\bar{\omega})-\Pi_4^{\beta\alpha_1\alpha_3\alpha_2}(\bar{\omega})=2\left(\Pi_5^{\beta\alpha_2\alpha_1\alpha_3}(\bar{\omega})-\Pi_5^{\beta\alpha_3\alpha_1\alpha_2}(\bar{\omega})\right)
\end{equation}
\section*{\hypertarget{B}{Appendix B}: Integral identities}

In the derivation of the expressions in Sections~\ref{Linearorder}-\ref{Thirdorder}, identities were used involving the parity and the derivatives of the integrals in the Eqs.~\ref{Pi1first},~\ref{Pi1second}-\ref{Pi2second},~\ref{Pi1third}-\ref{Pi6third}. These identities are helpful in manipulating general expressions and are summarized here. Their use is exemplified in Appendix~\hyperlink{C}{C}.\newline

The first set of identities concerns the derivatives of some $\Pi_i^{\alpha}$ integrals. In second order, 

\begin{equation}
    \frac{d}{d\bar{\omega}}\,\Pi_1^{\beta\alpha_1\alpha_2}(\bar{\omega})=\Pi_2^{\beta\alpha_1\alpha_2}(\bar{\omega})+\Pi_2^{\alpha_1\beta\alpha_2}(-\bar{\omega})+\Pi_B^{\beta\alpha_1\alpha_2}(\bar{\omega})
    \label{piderivativessecond}
\end{equation}
with 

\begin{equation}
    \Pi_B^{\beta\alpha_1\alpha_2}(\bar{\omega})\equiv\int\frac{d^d\mathbf{k}}{(2\pi)^d}\sum_{a,b}\frac{\Im\{\mathcal{A}^{\beta}_{ba}\,\mathcal{A}^{\alpha_1}_{ab}\}\,(\partial^{\alpha_2}\Delta f_{ba})}{\hbar\bar{\omega}-\Delta\epsilon_{ab}}
    \label{Pibsecond}
\end{equation}
which vanishes in the absence of a Fermi surface.

In third order,

\begin{align}
    \frac{d}{d\bar{\omega}}\,\Pi_1^{\beta\alpha_1\alpha_2\alpha_3}(\bar{\omega})&=\Pi_4^{\beta\alpha_1\alpha_2\alpha_3}(\bar{\omega})+\Pi_6^{\beta\alpha_3\alpha_1\alpha_2}(\bar{\omega})+\Pi_B^{\beta\alpha_1\alpha_2\alpha_3}(\bar{\omega})\\
    \frac{d}{d\bar{\omega}}\,\Pi_3^{\beta\alpha_1\alpha_2\alpha_3}(\bar{\omega})&=\Pi_1^{\beta\alpha_1\alpha_3\alpha_2}(\bar{\omega})-\Pi_1^{\alpha_1\beta\alpha_3\alpha_2}(-\bar{\omega})+\Pi_2^{\beta\alpha_1\alpha_2\alpha_3}(\bar{\omega})+\Pi_C^{\beta\alpha_1\alpha_2\alpha_3}(\bar{\omega})\nonumber\\
    &=\Pi_1^{\beta\alpha_1\alpha_2\alpha_3}(\bar{\omega})-\Pi_1^{\alpha_1\beta\alpha_2\alpha_3}(-\bar{\omega})+\Pi_2^{\beta\alpha_1\alpha_2\alpha_3}(\bar{\omega})+\Pi_C^{\beta\alpha_1\alpha_3\alpha_2}(\bar{\omega})
    \label{piderivativesthird}
\end{align}
with

\begin{align}
    \Pi_B^{\beta\alpha_1\alpha_2\alpha_3}(\bar{\omega})&\equiv\int\frac{d^d\mathbf{k}}{(2\pi)^d}\sum_{a,b}\frac{\Re\{\mathcal{A}^{\beta}_{ba}\,\mathcal{A}^{\alpha_1}_{ab;\alpha_2}\}\,(\partial^{\alpha_3}\Delta f_{ba})}{\hbar\bar{\omega}-\Delta\epsilon_{ab}}\\
    \Pi_C^{\beta\alpha_1\alpha_2\alpha_3}(\bar{\omega})&\equiv\int\frac{d^d\mathbf{k}}{(2\pi)^d}\sum_{a,b}\frac{\Re\{\mathcal{A}^{\beta}_{ba}\,\mathcal{A}^{\alpha_1}_{ab}\}\,(\partial^{\alpha_2}\Delta \epsilon_{ab})\,(\partial^{\alpha_3}\Delta f_{ba})}{\hbar\bar{\omega}-\Delta\epsilon_{ab}}
    \label{pibpicthird}
\end{align}
which vanish in the absence of a Fermi surface.

Additionally, from Eq.~\ref{piderivativesthird} it also follows,

\begin{equation}
    \Pi_1^{\beta\alpha_1\alpha_2\alpha_3}(\bar{\omega})-\Pi_1^{\beta\alpha_1\alpha_3\alpha_2}(\bar{\omega})-\Pi_1^{\alpha_1\beta\alpha_2\alpha_3}(-\bar{\omega})+\Pi_1^{\alpha_1\beta\alpha_3\alpha_2}(-\bar{\omega})=\Pi_C^{\beta\alpha_1\alpha_2\alpha_3}(\bar{\omega})-\Pi_C^{\beta\alpha_1\alpha_3\alpha_2}(\bar{\omega})
\end{equation}

The second set of identities, unlike the first, is reliant on time-reversal symmetry and consists on simply stating the parity of the integrals in Eqs.~\ref{Pi1first},~\ref{Pi1second}-\ref{Pi2second},~\ref{Pi1third}-\ref{Pi6third},

\begin{align}
    \Pi_1^{\beta\alpha_1}(-\bar{\omega})&=\Pi_1^{\beta\alpha_1}(\bar{\omega})\\
    \Pi_1^{\beta\alpha_1\alpha_2}(-\bar{\omega})&=\Pi_1^{\beta\alpha_1\alpha_2}(\bar{\omega})\\
    \Pi_2^{\beta\alpha_1\alpha_2}(-\bar{\omega})&=-\Pi_2^{\beta\alpha_1\alpha_2}(\bar{\omega})\\
    \Pi_1^{\beta\alpha_1\alpha_2\alpha_3}(-\bar{\omega})&=-\Pi_1^{\beta\alpha_1\alpha_2\alpha_3}(\bar{\omega})\\
    \Pi_2^{\beta\alpha_1\alpha_2\alpha_3}(-\bar{\omega})&=-\Pi_2^{\beta\alpha_1\alpha_2\alpha_3}(\bar{\omega})\\
    \Pi_3^{\beta\alpha_1\alpha_2\alpha_3}(-\bar{\omega})&=\Pi_3^{\beta\alpha_1\alpha_2\alpha_3}(\bar{\omega})\\
    \Pi_4^{\beta\alpha_1\alpha_2\alpha_3}(-\bar{\omega})&=\Pi_4^{\beta\alpha_1\alpha_2\alpha_3}(\bar{\omega})\\
    \Pi_5^{\beta\alpha_1\alpha_2\alpha_3}(-\bar{\omega})&=\Pi_5^{\beta\alpha_1\alpha_2\alpha_3}(\bar{\omega})\\
    \Pi_6^{\beta\alpha_1\alpha_2\alpha_3}(-\bar{\omega})&=\Pi_6^{\beta\alpha_1\alpha_2\alpha_3}(\bar{\omega})
    \label{parity}
\end{align}

Finally, there is a third set of identities that relate different tensor elements of a given integral,

\begin{align}
    F_A^{\beta\alpha_1}&=F_A^{\alpha_1\beta}\\
    \Pi_1^{\beta\alpha_1}(\bar{\omega})&=\Pi_1^{\alpha_1\beta}(\bar{\omega})\\
    F_B^{\beta\alpha_1\alpha_2}&=-F_B^{\alpha_1\beta\alpha_2}\\
    \Pi_1^{\beta\alpha_1\alpha_2}(\bar{\omega})&=-\Pi_1^{\alpha_1\beta\alpha_2}(\bar{\omega})\\
    \Pi_2^{\beta\alpha_1\alpha_2}(\bar{\omega})&=\Pi_2^{\beta\alpha_2\alpha_1}(\bar{\omega})\label{thisguy}\\
    F_A^{\beta\alpha_1\alpha_2\alpha_3}=F_A^{\alpha_1\beta\alpha_2\alpha_3}&=F_A^{\alpha_1\alpha_2\beta\alpha_3}=F_A^{\alpha_1\alpha_2\alpha_3\beta}\\
    \Pi_1^{\beta\alpha_1\alpha_2\alpha_3}(\bar{\omega})&=\Pi_1^{\beta\alpha_2\alpha_1\alpha_3}(\bar{\omega})\label{thisotherguy}\\
    \Pi_2^{\beta\alpha_1\alpha_2\alpha_3}(\bar{\omega})&=\Pi_2^{\alpha_1\beta\alpha_2\alpha_3}(\bar{\omega})=\Pi_2^{\beta\alpha_1\alpha_3\alpha_2}(\bar{\omega})\\
    \Pi_3^{\beta\alpha_1\alpha_2\alpha_3}(\bar{\omega})&=\Pi_3^{\alpha_1\beta\alpha_2\alpha_3}(\bar{\omega})=\Pi_3^{\beta\alpha_1\alpha_3\alpha_2}(\bar{\omega})\\
    \Pi_4^{\beta\alpha_1\alpha_2\alpha_3}(\bar{\omega})&=\Pi_4^{\beta\alpha_2\alpha_1\alpha_3}(\bar{\omega})\label{thisgirl}\\
    \Pi_5^{\beta\alpha_1\alpha_2\alpha_3}(\bar{\omega})=\Pi_5^{\alpha_1\beta\alpha_2\alpha_3}(\bar{\omega})&=\Pi_5^{\beta\alpha_1\alpha_3\alpha_2}(\bar{\omega})=\Pi_5^{\alpha_2\alpha_3\beta\alpha_1}(\bar{\omega})\\
    \Pi_6^{\beta\alpha_1\alpha_2\alpha_3}(\bar{\omega})=\Pi_6^{\alpha_1\beta\alpha_2\alpha_3}(\bar{\omega})&=\Pi_6^{\beta\alpha_1\alpha_3\alpha_2}(\bar{\omega})=\Pi_6^{\alpha_2\alpha_3\beta\alpha_1}(\bar{\omega})\label{thisothergirl}
\end{align}

These can be seen by inspection of the integrals, with Eqs.~\ref{thisguy},~\ref{thisotherguy},~\ref{thisgirl} and~\ref{thisothergirl} as a consequence of Eq.~\ref{identity2}.

These symmetries reduce the number of independent tensor elements that are required for a nonlinear conductivity calculation.
\section*{\hypertarget{C}{Appendix C}: Derivation of the second order conductivity}

As an illustration of the general scheme by which the expressions in Sections~\ref{Linearorder}-\ref{Thirdorder} were derived, the case of the second order conductivity is treated here in detail. The aim is to arrive at the results of Section~\ref{Secondorder}, starting with the general expression for a nonlinear conductivity. The derivation of third order conductivity, albeit considerably lengthier, follows along the same lines.

We start with Eq.~\ref{sigmaomegabar}, taken for the case $n=2$,

\begin{equation}
    \sigma^{\beta\alpha_1\alpha_2}(\bar{\omega}_1,\bar{\omega}_2)=\int\frac{d^d\mathbf{k}}{(2\pi)^d}\sum_{a,b}\frac{J^{\beta}_{\mathbf{k}ba}}{\hbar\bar{\omega}_1+\hbar\bar{\omega}_2-\Delta\epsilon_{\mathbf{k}ab}}\,\left[r^{\alpha_2},\frac{1}{\hbar\bar{\omega}_1-\Delta\epsilon_{\mathbf{k}}}\circ\left[r^{\alpha_1},\rho_0\right]\right]_{\mathbf{k}ab}
    \label{sigmaomegasecond}
\end{equation}

Substituting for the current matrix elements, Eq.~\ref{currentmatrixelements}, and using $\hat{r}^{\alpha}=i\hat{D}^{\alpha}$,

\begin{align}
    \sigma^{\beta\alpha_1\alpha_2}(\bar{\omega}_1,\bar{\omega}_2)&=-\frac{e^3}{\hbar}\,\int\frac{d^d\mathbf{k}}{(2\pi)^d}\sum_{a,b}\frac{\left[\hat{D}^{\beta},H_0\right]_{\mathbf{k}ba}}{\hbar\bar{\omega}_1+\hbar\bar{\omega}_2-\Delta\epsilon_{\mathbf{k}ab}}\,\left[r^{\alpha_2},\frac{1}{\hbar\bar{\omega}_1-\Delta\epsilon_{\mathbf{k}}}\circ\left[r^{\alpha_1},\rho_0\right]\right]_{\mathbf{k}ab}\nonumber\\
    &=\frac{e^3}{\hbar}\,\int\frac{d^d\mathbf{k}}{(2\pi)^d}\sum_{a}\frac{\partial^{\beta}\epsilon_{\mathbf{k}a}}{\hbar\bar{\omega}_1+\hbar\bar{\omega}_2}\,\left[D^{\alpha_2},\frac{1}{\hbar\bar{\omega}_1-\Delta\epsilon_{\mathbf{k}}}\circ\left[D^{\alpha_1},\rho_0\right]\right]_{\mathbf{k}aa}\nonumber\\
    &-\frac{i\,e^3}{\hbar}\,\int\frac{d^d\mathbf{k}}{(2\pi)^d}\sum_{a\neq b}\frac{\mathcal{A}^{\beta}_{\mathbf{k}ba}\,\Delta\epsilon_{\mathbf{k}ab}}{\hbar\bar{\omega}_1+\hbar\bar{\omega}_2-\Delta\epsilon_{\mathbf{k}ab}}\,\left[D^{\alpha_2},\frac{1}{\hbar\bar{\omega}_1-\Delta\epsilon_{\mathbf{k}}}\circ\left[D^{\alpha_1},\rho_0\right]\right]_{\mathbf{k}ab}
    \label{sigmaomegasecondv2}
\end{align}
where diagonal and off-diagonal current matrix elements were separated.

To expand the condensed expression in Eq.~\ref{sigmaomegasecondv2} into more explicit formulae, Eq.~\ref{dcommutator} is applied iteratively, starting with

\begin{equation}
    [\hat{D}^{\alpha},\hat{\rho_0}]_{\mathbf{k}ab}=\delta_{ab}\,\partial^{\alpha}f_{\mathbf{k}a}-i\mathcal{A}^{\alpha}_{\mathbf{k}ab}\Delta f_{\mathbf{k}ba}
\end{equation}
where $\delta_{ab}$ is the Kronecker delta.

In the intermediate steps, the definition of the Hadamard product is used. For instance,

\begin{equation}
    \left(\frac{1}{\hbar\bar{\omega}_1-\Delta\epsilon_{\mathbf{k}}}\circ\left[D^{\alpha_1},\rho_0\right]\right)_{\mathbf{k}aa}=\frac{\left[D^{\alpha_1},\rho_0\right]_{\mathbf{k}aa}}{\hbar\bar{\omega}_1}=\frac{\partial^{\alpha_1} f_{\mathbf{k}a}}{\hbar\bar{\omega}_1}
\end{equation}

\begin{equation}
    \left(\frac{1}{\hbar\bar{\omega}_1-\Delta\epsilon_{\mathbf{k}}}\circ\left[D^{\alpha_1},\rho_0\right]\right)_{\mathbf{k}ab}=\frac{\left[D^{\alpha_1},\rho_0\right]_{\mathbf{k}ab}}{\hbar\bar{\omega}_1-\Delta\epsilon_{\mathbf{k}ab}}=-\frac{i\,\mathcal{A}^{\alpha_1}_{\mathbf{k}ab}\Delta f_{\mathbf{k}ba}}{\hbar\bar{\omega}_1-\Delta\epsilon_{\mathbf{k}ab}}
\end{equation}
with $a\neq b$ in the last equation.

Expanding the first commutator in Eq.~\ref{sigmaomegasecondv2},

\begin{align}
    \left[D^{\alpha_2},\frac{1}{\hbar\bar{\omega}_1-\Delta\epsilon_{\mathbf{k}}}\circ\left[D^{\alpha_1},\rho_0\right]\right]_{\mathbf{k}aa}&=\frac{\partial^{\alpha_2}\partial^{\alpha_1}f_{\mathbf{k}a}}{\hbar\,\bar{\omega}_1}+\sum_{c}\frac{\mathcal{A}^{\alpha_2}_{\mathbf{k}ac}\mathcal{A}^{\alpha_1}_{\mathbf{k}ca}\Delta f_{\mathbf{k}ca}}{\hbar\bar{\omega}_1+\Delta\epsilon_{\mathbf{k}ac}}+\sum_{c}\frac{\mathcal{A}^{\alpha_2}_{\mathbf{k}ca}\mathcal{A}^{\alpha_1}_{\mathbf{k}ac}\Delta f_{\mathbf{k}ca}}{\hbar\bar{\omega}_1-\Delta\epsilon_{\mathbf{k}ac}}
    \label{diagonalcommutator}
\end{align}
while the second commutator gives, for $a\neq b$,

\begin{align}
    \left[D^{\alpha_2},\frac{1}{\hbar\bar{\omega}_1-\Delta\epsilon_{\mathbf{k}}}\circ\left[D^{\alpha_1},\rho_0\right]\right]_{\mathbf{k}ab}&=-\frac{i}{\hbar\bar{\omega}_1}\,\mathcal{A}^{\alpha_2}_{\mathbf{k}ab}\,(\partial^{\alpha_1}\,\Delta f_{\mathbf{k}ba})-\frac{i\,\mathcal{A}^{\alpha_1}_{\mathbf{k}ab;\alpha_2}\,\Delta f_{\mathbf{k}ba}}{\hbar\bar{\omega}_1-\Delta\epsilon_{\mathbf{k}ab}}\nonumber\\
    &-\frac{i\,\mathcal{A}^{\alpha_1}_{\mathbf{k}ab}\,(\partial^{\alpha_2}\Delta f_{\mathbf{k}ba})}{\hbar\bar{\omega}_1-\Delta\epsilon_{\mathbf{k}ab}}-\frac{i\,\mathcal{A}^{\alpha_1}_{\mathbf{k}ab}\,(\partial^{\alpha_2}\Delta\epsilon_{\mathbf{k}ab})\,\Delta f_{\mathbf{k}ba}}{(\hbar\bar{\omega}_1-\Delta\epsilon_{\mathbf{k}ab})^2}
    \label{offdiagonalcommutator}
\end{align}

Replacing Eqs.~\ref{diagonalcommutator} and~\ref{offdiagonalcommutator} in Eq.~\ref{sigmaomegasecondv2},

\begin{align}
    \sigma^{\beta\alpha_1\alpha_2}(\bar{\omega}_1,\bar{\omega}_2)&=\frac{e^3}{\hbar}\,\frac{1}{\hbar\bar{\omega}_1\,(\hbar\bar{\omega}_1+\hbar\bar{\omega}_2)}\int\frac{d^d\mathbf{k}}{(2\pi)^d}\sum_a\partial^{\beta}\epsilon_{\mathbf{k}a}\,\partial^{\alpha_1}\partial^{\alpha_2}f_{\mathbf{k}a}\nonumber\\
    &+\frac{e^3}{\hbar}\,\frac{1}{\hbar\bar{\omega}_1+\hbar\bar{\omega}_2}\int\frac{d^d\mathbf{k}}{(2\pi)^d}\sum_{a,b}\frac{\mathcal{A}^{\alpha_2}_{\mathbf{k}ba}\mathcal{A}^{\alpha_1}_{\mathbf{k}ab}(\partial^{\beta}\Delta\epsilon_{\mathbf{k}ab})\Delta f_{\mathbf{k}ba}}{\hbar\bar{\omega}_1-\Delta\epsilon_{\mathbf{k}ab}}\nonumber\\
    &-\frac{e^3}{\hbar}\,\frac{1}{\hbar\bar{\omega}_1}\int\frac{d^d\mathbf{k}}{(2\pi)^d}\sum_{a,b}\frac{\mathcal{A}^{\beta}_{\mathbf{k}ba}\mathcal{A}^{\alpha_2}_{\mathbf{k}ab}\Delta\epsilon_{\mathbf{k}ab}(\partial^{\alpha_1}\Delta f_{\mathbf{k}ba})}{\hbar\bar{\omega}_1+\hbar\bar{\omega}_2-\Delta\epsilon_{\mathbf{k}ab}}\nonumber\\
    &-\frac{e^3}{\hbar}\,\int\frac{d^d\mathbf{k}}{(2\pi)^d}\sum_{a,b}\frac{\mathcal{A}^{\beta}_{\mathbf{k}ba}\mathcal{A}^{\alpha_1}_{\mathbf{k}ab}\Delta\epsilon_{\mathbf{k}ab}(\partial^{\alpha_2}\Delta f_{\mathbf{k}ba})}{(\hbar\bar{\omega}_1-\Delta\epsilon_{\mathbf{k}ab})(\hbar\bar{\omega}_1+\hbar\bar{\omega}_2-\Delta\epsilon_{\mathbf{k}ab})}\nonumber\\    
    &-\frac{e^3}{\hbar}\,\int\frac{d^d\mathbf{k}}{(2\pi)^d}\sum_{a,b}\frac{\mathcal{A}^{\beta}_{\mathbf{k}ba}\mathcal{A}^{\alpha_1}_{\mathbf{k}ab;\alpha_2}\,\Delta f_{\mathbf{k}ba}}{(\hbar\bar{\omega}_1-\Delta\epsilon_{\mathbf{k}ab})(\hbar\bar{\omega}_1+\hbar\bar{\omega}_2-\Delta\epsilon_{\mathbf{k}ab})}\nonumber\\
    &-\frac{e^3}{\hbar}\,\int\frac{d^d\mathbf{k}}{(2\pi)^d}\sum_{a,b}\frac{\mathcal{A}^{\beta}_{\mathbf{k}ba}\mathcal{A}^{\alpha_1}_{\mathbf{k}ab}\,(\partial^{\alpha_2}\Delta\epsilon_{ab})\,\Delta\epsilon_{\mathbf{k}ab}\,\Delta f_{\mathbf{k}ba}}{(\hbar\bar{\omega}_1-\Delta\epsilon_{\mathbf{k}ab})^2(\hbar\bar{\omega}_1+\hbar\bar{\omega}_2-\Delta\epsilon_{\mathbf{k}ab})}  
    \label{sigmaomegasecondv3}
\end{align}

Imposing time-reversal symmetry, we have $\epsilon_{-\mathbf{k}a}=\epsilon_{\mathbf{k}a}$ and  $\mathcal{A}_{-\mathbf{k}ab}=\mathcal{A}_{\mathbf{k}ba}$. Since it is only a function of energy, we have also $f_{-\mathbf{k}a}=f_{\mathbf{k}a}$. As a result, the first term in Eq.~\ref{sigmaomegasecondv3} vanishes due to the odd number of derivatives in $\mathbf{k}$, while the other terms are shown to have purely imaginary numerators by combining the contributions at $\mathbf{k}$ and $-\mathbf{k}$. For instance,

\begin{align}
    \int\frac{d^d\mathbf{k}}{(2\pi)^d}&\sum_{a,b}\frac{\mathcal{A}^{\alpha_2}_{\mathbf{k}ba}\mathcal{A}^{\alpha_1}_{\mathbf{k}ab}(\partial^{\beta}\Delta\epsilon_{\mathbf{k}ab})\Delta f_{\mathbf{k}ba}}{\hbar\bar{\omega}_1-\Delta\epsilon_{\mathbf{k}ab}}\nonumber\\
    &=\frac{1}{2}\int\frac{d^d\mathbf{k}}{(2\pi)^d}\sum_{a,b}\frac{\mathcal{A}^{\alpha_2}_{\mathbf{k}ba}\mathcal{A}^{\alpha_1}_{\mathbf{k}ab}(\partial^{\beta}\Delta\epsilon_{\mathbf{k}ab})\Delta f_{\mathbf{k}ba}}{\hbar\bar{\omega}_1-\Delta\epsilon_{\mathbf{k}ab}}+\frac{1}{2}\int\frac{d^d\mathbf{k}}{(2\pi)^d}\sum_{a,b}\frac{\mathcal{A}^{\alpha_2}_{\mathbf{k}ba}\mathcal{A}^{\alpha_1}_{\mathbf{k}ab}(\partial^{\beta}\Delta\epsilon_{\mathbf{k}ab})\Delta f_{\mathbf{k}ba}}{\hbar\bar{\omega}_1-\Delta\epsilon_{\mathbf{k}ab}}\nonumber\\
    &=\frac{1}{2}\int\frac{d^d\mathbf{k}}{(2\pi)^d}\sum_{a,b}\frac{\mathcal{A}^{\alpha_2}_{\mathbf{k}ba}\mathcal{A}^{\alpha_1}_{\mathbf{k}ab}(\partial^{\beta}\Delta\epsilon_{\mathbf{k}ab})\Delta f_{\mathbf{k}ba}}{\hbar\bar{\omega}_1-\Delta\epsilon_{\mathbf{k}ab}}+\frac{1}{2}\int\frac{d^d\mathbf{k}}{(2\pi)^d}\sum_{a,b}\frac{\mathcal{A}^{\alpha_2}_{-\mathbf{k}ba}\mathcal{A}^{\alpha_1}_{-\mathbf{k}ab}(-\partial^{\beta}\Delta\epsilon_{-\mathbf{k}ab})\Delta f_{-\mathbf{k}ba}}{\hbar\bar{\omega}_1-\Delta\epsilon_{-\mathbf{k}ab}}\nonumber\\
    &=\frac{1}{2}\int\frac{d^d\mathbf{k}}{(2\pi)^d}\sum_{a,b}\frac{\mathcal{A}^{\alpha_2}_{\mathbf{k}ba}\mathcal{A}^{\alpha_1}_{\mathbf{k}ab}(\partial^{\beta}\Delta\epsilon_{\mathbf{k}ab})\Delta f_{\mathbf{k}ba}}{\hbar\bar{\omega}_1-\Delta\epsilon_{\mathbf{k}ab}}-\frac{1}{2}\int\frac{d^d\mathbf{k}}{(2\pi)^d}\sum_{a,b}\frac{\mathcal{A}^{\alpha_2}_{\mathbf{k}ab}\mathcal{A}^{\alpha_1}_{\mathbf{k}ba}(\partial^{\beta}\Delta\epsilon_{\mathbf{k}ab})\Delta f_{\mathbf{k}ba}}{\hbar\bar{\omega}_1-\Delta\epsilon_{\mathbf{k}ab}}\nonumber\\
    &=\int\frac{d^d\mathbf{k}}{(2\pi)^d}\sum_{a,b}\frac{i\Im\{\mathcal{A}^{\alpha_2}_{\mathbf{k}ba}\,\mathcal{A}^{\alpha_1}_{\mathbf{k}ab}\}(\partial^{\beta}\Delta\epsilon_{\mathbf{k}ab})\Delta f_{\mathbf{k}ba}}{\hbar\bar{\omega}_1-\Delta\epsilon_{\mathbf{k}ab}}
\end{align}
with the use of Eq.~\ref{antisymmetriccomb}. This procedure is to be repeated for the remaining terms in Eq.~\ref{sigmaomegasecondv3}.

The next step is a partial fraction decomposition,

\begin{align}
    &\sigma^{\beta\alpha_1\alpha_2}(\bar{\omega}_1,\bar{\omega}_2)=\frac{i e^3}{\hbar}\,\frac{1}{\hbar\bar{\omega}_1+\hbar\bar{\omega}_2}\int\frac{d^d\mathbf{k}}{(2\pi)^d}\sum_{a,b}\frac{\Im\{\mathcal{A}^{\alpha_2}_{\mathbf{k}ba}\,\mathcal{A}^{\alpha_1}_{\mathbf{k}ab}\}(\partial^{\beta}\Delta\epsilon_{\mathbf{k}ab})\Delta f_{\mathbf{k}ba}}{\hbar\bar{\omega}_1-\Delta\epsilon_{\mathbf{k}ab}}\nonumber\\
    &-\frac{i e^3}{\hbar}\,\frac{1}{\hbar\bar{\omega}_1}\int\frac{d^d\mathbf{k}}{(2\pi)^d}\sum_{a,b}\Im\{\mathcal{A}^{\beta}_{\mathbf{k}ba}\,\mathcal{A}^{\alpha_2}_{\mathbf{k}ab}\}(\partial^{\alpha_1}\Delta f_{\mathbf{k}ba})\left(\frac{\hbar\bar{\omega}_1+\hbar\bar{\omega}_2}{\hbar\bar{\omega}_1+\hbar\bar{\omega}_2-\Delta\epsilon_{\mathbf{k}ab}}-1\right)\nonumber\\
    &-\frac{i e^3}{\hbar}\,\int\frac{d^d\mathbf{k}}{(2\pi)^d}\sum_{a,b}\Im\{\mathcal{A}^{\beta}_{\mathbf{k}ba}\,\mathcal{A}^{\alpha_1}_{\mathbf{k}ab}\}(\partial^{\alpha_2}\Delta f_{\mathbf{k}ba})\left(\frac{\hbar\bar{\omega}_1}{\hbar\bar{\omega}_2}\frac{1}{\hbar\bar{\omega}_1-\Delta\epsilon_{\mathbf{k}ab}}-\frac{\hbar\bar{\omega}_1+\hbar\bar{\omega}_2}{\hbar\bar{\omega}_2}\frac{1}{\hbar\bar{\omega}_1+\hbar\bar{\omega}_2-\Delta\epsilon_{\mathbf{k}ab}}\right)\nonumber\\    
    &-\frac{i e^3}{\hbar}\,\int\frac{d^d\mathbf{k}}{(2\pi)^d}\sum_{a,b}\Im\{\mathcal{A}^{\beta}_{\mathbf{k}ba}\,\mathcal{A}^{\alpha_1}_{\mathbf{k}ab;\alpha_2}\}\,\Delta f_{\mathbf{k}ba}\left(\frac{\hbar\bar{\omega}_1}{\hbar\bar{\omega}_2}\frac{1}{\hbar\bar{\omega}_1-\Delta\epsilon_{\mathbf{k}ab}}-\frac{\hbar\bar{\omega}_1+\hbar\bar{\omega}_2}{\hbar\bar{\omega}_2}\frac{1}{\hbar\bar{\omega}_1+\hbar\bar{\omega}_2-\Delta\epsilon_{\mathbf{k}ab}}\right)\nonumber\\
    &-\frac{i e^3}{\hbar}\,\int\frac{d^d\mathbf{k}}{(2\pi)^d}\sum_{a,b}\Im\{\mathcal{A}^{\beta}_{\mathbf{k}ba}\,\mathcal{A}^{\alpha_1}_{\mathbf{k}ab}\}\,(\partial^{\alpha_2}\Delta\epsilon_{ab})\,\Delta f_{\mathbf{k}ba}\nonumber\\
    &\qquad\qquad\qquad\times\left(\frac{\hbar\bar{\omega}_1}{\hbar\bar{\omega}_2}\frac{1}{(\hbar\bar{\omega}_1-\Delta\epsilon_{\mathbf{k}ab})^2}-\frac{\hbar\bar{\omega}_1+\hbar\bar{\omega}_2}{(\hbar\bar{\omega}_2)^2}\frac{1}{\hbar\bar{\omega}_1-\Delta\epsilon_{\mathbf{k}ab}}+\frac{\hbar\bar{\omega}_1+\hbar\bar{\omega}_2}{(\hbar\bar{\omega}_2)^2}\frac{1}{\hbar\bar{\omega}_1+\hbar\bar{\omega}_2-\Delta\epsilon_{\mathbf{k}ab}}\right)
    \label{partialfractiondecomposition}
\end{align}

The second term in the second line of Eq.~\ref{partialfractiondecomposition} can be singled out and identified as a Fermi surface contribution,

\begin{align}
    &\frac{i e^3}{\hbar}\,\frac{1}{\hbar\bar{\omega}_1}\int\frac{d^d\mathbf{k}}{(2\pi)^d}\sum_{a,b}\Im\{\mathcal{A}^{\beta}_{\mathbf{k}ba}\,\mathcal{A}^{\alpha_2}_{\mathbf{k}ab}\}(\partial^{\alpha_1}\Delta f_{\mathbf{k}ba})=\frac{i e^3}{\hbar}\,\frac{1}{\hbar\bar{\omega}_1}\int\frac{d^d\mathbf{k}}{(2\pi)^d}\sum_{a,b}\frac{1}{2 i}\left(\mathcal{A}^{\beta}_{\mathbf{k}ba}\mathcal{A}^{\alpha_2}_{\mathbf{k}ab}-\mathcal{A}^{\beta}_{\mathbf{k}ab}\mathcal{A}^{\alpha_2}_{\mathbf{k}ba}\right)(\partial^{\alpha_1}\Delta f_{\mathbf{k}ba})\nonumber\\
    &=\frac{i e^3}{\hbar}\,\frac{1}{\hbar\bar{\omega}_1}\int\frac{d^d\mathbf{k}}{(2\pi)^d}\sum_{b}(-i)\left[\mathcal{A}^{\beta},\mathcal{A}^{\alpha_2}\right]_{\mathbf{k}bb}(\partial^{\alpha_1}f_{\mathbf{k}b})=\frac{i e^3}{\hbar}\,\frac{1}{\hbar\bar{\omega}_1}\int\frac{d^d\mathbf{k}}{(2\pi)^d}\sum_{b}\mathcal{F}^{\alpha_2\beta}_{\mathbf{k}b}(\partial^{\alpha_1}f_{\mathbf{k}b})=\frac{i e^3}{\hbar}\,\frac{1}{\hbar\bar{\omega}_1}\,F_{B}^{\alpha_1\alpha_2\beta}
\end{align}
with the use of Eqs.~\ref{antisymmetriccomb} and~\ref{berrycurvatureandcommutator}.

Identifying the remaining integrals with the definitions in Eqs.~\ref{Pi1second}-\ref{Pi2second} and~\ref{Pibsecond},

\begin{align}
    -\frac{\hbar}{i\,e^3}\,\sigma^{\beta\alpha_1\alpha_2}(\bar{\omega}_1,\bar{\omega}_2)=&+\frac{\bar{\omega}_1}{\bar{\omega}_2}\,\Pi_B^{\beta\alpha_1\alpha_2}(\bar{\omega}_1)-\frac{\bar{\omega}_1+\bar{\omega}_2}{\bar{\omega}_2}\,\Pi_B^{\beta\alpha_1\alpha_2}(\bar{\omega}_1+\bar{\omega}_2)+\frac{\bar{\omega}_1+\bar{\omega}_2}{\bar{\omega}_2}\,\Pi_B^{\beta\alpha_2\alpha_1}(\bar{\omega}_1+\bar{\omega}_2)\nonumber\\
    &-\frac{1}{\bar{\omega}_1+\bar{\omega}_2}\,\Pi_1^{\alpha_2\alpha_1\beta}(\bar{\omega}_1)-\frac{\bar{\omega}_1+\bar{\omega}_2}{\bar{\omega}_2^2}\,\Pi_1^{\beta\alpha_1\alpha_2}(\bar{\omega}_1)-\frac{\bar{\omega}_1}{\bar{\omega}_2}\,\frac{d}{d\bar{\omega}_1}\Pi_1^{\beta\alpha_1\alpha_2}(\bar{\omega}_1)+\frac{\bar{\omega}_1}{\bar{\omega}_2}\,\Pi_2^{\beta\alpha_1\alpha_2}(\bar{\omega}_1)\nonumber\\
    &+\frac{\bar{\omega}_1+\bar{\omega}_2}{\bar{\omega}_2^2}\,\Pi_1^{\beta\alpha_1\alpha_2}(\bar{\omega}_1+\bar{\omega}_2)-\frac{\bar{\omega}_1+\bar{\omega}_2}{\bar{\omega}_2}\,\Pi_2^{\beta\alpha_1\alpha_2}(\bar{\omega}_1+\bar{\omega}_2)
    \label{sigmaomegasecondintegrals}
\end{align}
and using Eq.~\ref{piderivativessecond} to replace the derivative, we obtain

\begin{align}
    \frac{\hbar}{i\,e^3}\,\sigma^{\beta\alpha_1\alpha_2}(\bar{\omega}_1,\bar{\omega}_2)=&\frac{1}{\bar{\omega}_1+\bar{\omega}_2}\,\Pi_1^{\alpha_2\alpha_1\beta}(\bar{\omega}_1)+\frac{\bar{\omega}_1+\bar{\omega}_2}{\bar{\omega}_2^2}\,\Pi_1^{\beta\alpha_1\alpha_2}(\bar{\omega}_1)-\frac{\bar{\omega}_1}{\bar{\omega}_2}\,\Pi_2^{\alpha_1\beta\alpha_2}(\bar{\omega}_1)\nonumber\\
    &-\frac{\bar{\omega}_1+\bar{\omega}_2}{\bar{\omega}_2^2}\,\Pi_1^{\beta\alpha_1\alpha_2}(\bar{\omega}_1+\bar{\omega}_2)+\frac{\bar{\omega}_1+\bar{\omega}_2}{\bar{\omega}_2}\,\Pi_2^{\beta\alpha_1\alpha_2}(\bar{\omega}_1+\bar{\omega}_2)
    \label{sigmaomegasecondfinalversion}
\end{align}
in agreement with Eqs.~\ref{sigma2}-\ref{Pi2second}.
\section*{\hypertarget{D}{Appendix D}: Integral evaluation for monolayer graphene}

The evaluation of the integrals in Eqs.~\ref{FAfirst}-\ref{Pi1first},~\ref{FBsecond}-\ref{Pi2second} and~\ref{FAthird}-\ref{Pi6third} for monolayer graphene near the Dirac point is relatively simple. Eqs~\ref{conductionbandgraphene} and~\ref{valencebandgraphene} for the band energies and Eq.~\ref{nonabelianberryconnection} for the non-Abelian Berry connection are substituted and their derivatives (or generalized derivatives) are computed. Taking the relaxation-free limit and considering $T=0$, the $\mathcal{I}$ integrals that define the real part can be easily computed by converting the integral over the FBZ to polar coordinates and using the Dirac delta function in Eq.~\ref{I}.

It is important to note that the integrals must respect the same symmetries, as far as the tensor indices $\beta\alpha_1\dots\alpha_n$ are concerned, as the conductivities. This considerably reduces the number of integrals to compute, since, for graphene,

\begin{align}
    \sigma^{xy}(\omega)&=\sigma^{yx}(\omega)=0\\
    \sigma^{xx}(\omega)&=\sigma^{yy}(\omega)
\end{align}
at linear order. The second order is absent and at third order many tensor components vanish,

\begin{align}
    &\sigma^{xxxy}(\omega_1,\omega_2,\omega_3)=\sigma^{xxyx}(\omega_1,\omega_2,\omega_3)=\sigma^{xyxx}(\omega_1,\omega_2,\omega_3)=\sigma^{yxxx}(\omega_1,\omega_2,\omega_3)\nonumber\\
    &=\sigma^{yyyx}(\omega_1,\omega_2,\omega_3)=\sigma^{yyxy}(\omega_1,\omega_2,\omega_3)=\sigma^{yxyy}(\omega_1,\omega_2,\omega_3)=\sigma^{xyyy}(\omega_1,\omega_2,\omega_3)=0
\end{align}
while the others are interrelated,

\begin{align}
    &\sigma^{xxxx}(\omega_1,\omega_2,\omega_3)=\sigma^{yyyy}(\omega_1,\omega_2,\omega_3)\qquad\sigma^{xxyy}(\omega_1,\omega_2,\omega_3)=\sigma^{yyxx}(\omega_1,\omega_2,\omega_3)\\
    &\sigma^{xyxy}(\omega_1,\omega_2,\omega_3)=\sigma^{yxyx}(\omega_1,\omega_2,\omega_3)\qquad\sigma^{xyyx}(\omega_1,\omega_2,\omega_3)=\sigma^{yxxy}(\omega_1,\omega_2,\omega_3)\\
    &\sigma^{xxxx}(\omega_1,\omega_2,\omega_3)=\sigma^{xxyy}(\omega_1,\omega_2,\omega_3)+\sigma^{xyxy}(\omega_1,\omega_2,\omega_3)+\sigma^{xyyx}(\omega_1,\omega_2,\omega_3)
\end{align}
with only three independent components remaining: $\sigma^{xxyy}$, $\sigma^{xyxy}$ and $\sigma^{xyyx}$. This leaves us with one and three tensor elements to evaluate for each integral at linear and third order, respectively.

For linear order,

\begin{equation}
    \mathcal{I}_1^{xx}(\omega)=\frac{1}{4\pi\hbar\omega}\,\Theta(\hbar|\omega|-2|\mu|)
\end{equation}
whose Hilbert transform $\mathcal{H}_1$,

\begin{equation}
    \mathcal{H}_1^{xx}(\omega)=-\frac{1}{\pi}\dashint\frac{\mathcal{I}_1^{xx}(\omega')}{\omega'-\omega}d\omega'=\frac{1}{4\pi^2\hbar\omega}\,\ln\left|\frac{\hbar\omega-2|\mu|}{\hbar\omega+2|\mu|}\right|
    \label{Hilberttransformlinear}
\end{equation}

There is also the Fermi surface contribution $F_A$,
\begin{equation}
    F_A^{xx}=-\frac{|\mu|}{\pi}
\end{equation}

The case of the third order conductivity is entirely analogous, but there are, of course, more tensor components to assess now,

\begin{align}
    \mathcal{I}_1^{xxyy}(\omega)&=\mathcal{I}_1^{xyxy}(\omega)=-\frac{\hbar^2v_F^2}{2\pi(\hbar\omega)^2}\,\Theta(\hbar|\omega|-2|\mu|)\\
    \mathcal{I}_1^{xyyx}(\omega)&=\frac{\hbar^2v_F^2}{2\pi(\hbar\omega)^2}\,\Theta(\hbar|\omega|-2|\mu|)\\
    \mathcal{I}_2^{xxyy}(\omega)&=\mathcal{I}_2^{xyxy}(\omega)=\mathcal{I}_2^{xyyx}(\omega)=\frac{\hbar^2v_F^2}{4\pi(\hbar\omega)^2}\,\Theta(\hbar|\omega|-2|\mu|)\\
    \mathcal{I}_3^{xxyy}(\omega)&=\frac{3\hbar^2v_F^2}{4\pi\hbar\omega}\,\Theta(\hbar|\omega|-2|\mu|)\\
    \mathcal{I}_3^{xyxy}(\omega)&=\mathcal{I}_3^{xyyx}(\omega)=-\frac{\hbar^2v_F^2}{4\pi\hbar\omega}\,\Theta(\hbar|\omega|-2|\mu|)\\
    \mathcal{I}_4^{xxyy}(\omega)&=\mathcal{I}_4^{xyxy}(\omega)=\mathcal{I}_4^{xyyx}(\omega)=0\\
    \mathcal{I}_5^{xxyy}(\omega)&=\mathcal{I}_5^{xyxy}(\omega)=\mathcal{I}_5^{xyyx}(\omega)=\frac{\hbar^2v_F^2}{16\pi(\hbar\omega)^3}\,\Theta(\hbar|\omega|-2|\mu|)\\  
    \mathcal{I}_6^{xxyy}(\omega)&=-\frac{\hbar^2v_F^2}{\pi(\hbar\omega)^3}\,\Theta(\hbar|\omega|-2|\mu|)\\
    \mathcal{I}_6^{xyxy}(\omega)&=\mathcal{I}_6^{xyyx}(\omega)=\frac{\hbar^2v_F^2}{\pi(\hbar\omega)^3}\,\Theta(\hbar|\omega|-2|\mu|)
\end{align}
and the respective Hilbert transforms,

\begin{align}
    \mathcal{H}_1^{xxyy}(\omega)&=\mathcal{H}_1^{xyxy}(\omega)=-\frac{\hbar^2v_F^2}{2\pi^2(\hbar\omega)^2}\,\ln\left|\frac{\hbar\omega-2|\mu|}{\hbar\omega+2|\mu|}\right|-\frac{\hbar^2v_F^2}{2\pi^2|\mu|\hbar\omega}\\
    \mathcal{H}_1^{xyyx}(\omega)&=\frac{\hbar^2v_F^2}{2\pi^2(\hbar\omega)^2}\,\ln\left|\frac{\hbar\omega-2|\mu|}{\hbar\omega+2|\mu|}\right|+\frac{\hbar^2v_F^2}{2\pi^2|\mu|\hbar\omega}\\
    \mathcal{H}_2^{xxyy}(\omega)&=\mathcal{H}_2^{xyxy}(\omega)=\mathcal{H}_2^{xyyx}(\omega)=\frac{\hbar^2v_F^2}{4\pi^2(\hbar\omega)^2}\,\ln\left|\frac{\hbar\omega-2|\mu|}{\hbar\omega+2|\mu|}\right|+\frac{\hbar^2v_F^2}{4\pi^2|\mu|\hbar\omega}\\
    \mathcal{H}_3^{xxyy}(\omega)&=\frac{3\hbar^2v_F^2}{4\pi^2\hbar\omega}\,\ln\left|\frac{\hbar\omega-2|\mu|}{\hbar\omega+2|\mu|}\right|\\
    \mathcal{H}_3^{xyxy}(\omega)&=\mathcal{H}_3^{xyyx}(\omega)=-\frac{\hbar^2v_F^2}{4\pi^2\hbar\omega}\,\ln\left|\frac{\hbar\omega-2|\mu|}{\hbar\omega+2|\mu|}\right|\\
    \mathcal{H}_4^{xxyy}(\omega)&=\mathcal{H}_4^{xyxy}(\omega)=\mathcal{H}_4^{xyyx}(\omega)=0\\
    \mathcal{H}_5^{xxyy}(\omega)&=\mathcal{H}_5^{xyxy}(\omega)=\mathcal{H}_5^{xyyx}(\omega)=\frac{\hbar^2v_F^2}{16\pi^2(\hbar\omega)^3}\,\ln\left|\frac{\hbar\omega-2|\mu|}{\hbar\omega+2|\mu|}\right|+\frac{\hbar^2v_F^2}{16\pi^2|\mu|(\hbar\omega)^2}\\
    \mathcal{H}_6^{xxyy}(\omega)&=-\frac{\hbar^2v_F^2}{\pi^2(\hbar\omega)^3}\,\ln\left|\frac{\hbar\omega-2|\mu|}{\hbar\omega+2|\mu|}\right|-\frac{\hbar^2v_F^2}{\pi^2|\mu|(\hbar\omega)^2}\\
    \mathcal{H}_6^{xyxy}(\omega)&=\mathcal{H}_6^{xyyx}(\omega)=\frac{\hbar^2v_F^2}{\pi^2(\hbar\omega)^3}\,\ln\left|\frac{\hbar\omega-2|\mu|}{\hbar\omega+2|\mu|}\right|+\frac{\hbar^2v_F^2}{\pi^2|\mu|(\hbar\omega)^2}
\end{align}

For the passage from the $\mathcal{I}$ to the $\mathcal{H}$ integrals, it is useful to hold in mind the following result for $n>1$,

\begin{equation}
    -\frac{1}{\pi}\dashint\frac{\Theta(\omega'-\Delta_{eff})}{(\omega')^{n}(\omega-\omega')}d\omega'=\frac{1}{\pi\omega^n}\,\ln\left|\frac{\hbar\omega-\Delta_{eff}}{\hbar\omega+\Delta_{eff}}\right|+\frac{2}{\pi}\sum_{k=1}^{n-1}\frac{1}{(2k-1)\,\Delta_{eff}^{2k-1}\,(\hbar\omega)^{n+1-2k}}
    \label{Hilberttransform}
\end{equation}

Finally, there is the Fermi surface contribution,

\begin{align}
    F_A^{xxyy}=F_A^{xyxy}=F_A^{xyyx}=\frac{\hbar^2v_F^2}{4\pi|\mu|}
\end{align}

Replacement of the integrals computed here in the Eqs.~\ref{sigma3F}-\ref{sigma33} of Section~\ref{Resonancebasedanalysis} reproduces the expression for the third order conductivity derived by Cheng et al.~\cite{Cheng2014ThirdGraphene}.

\bibliography{references}

\end{document}